\documentclass[twocolumn,tighten,times]{aastex631}
\usepackage{hyperref}
\usepackage{amsmath}

\mathchardef\hyph="2D

\graphicspath{{./}{figures/}}

\submitjournal{ApJ}

\shorttitle{Line Locked Systems in Active Galaxies}
\shortauthors{Lewis \& Chelouche}
\begin{document}
\everymath{\displaystyle}

\title{\textbf{On The Fine Tuning and Physical Origin of Line-Locked Absorption Systems in Active Galaxies}}

\author[0000-0002-9854-1432]{Tiffany R. Lewis}
\altaffiliation{Zuckerman Fellow, NPP Fellow}
\affil{Haifa Center for Theoretical Physics and Astrophysics (HCTPA), University of Haifa, Haifa 3498838, Israel}
\affil{Astrophysics Science Division, NASA Goddard Space Flight Center, Greenbelt, MD 20771, USA}

\author[0000-0002-4830-7787]{Doron Chelouche}
\affil{Haifa Center for Theoretical Physics and Astrophysics (HCTPA), University of Haifa, Haifa 3498838, Israel}
\affil{Department of Physics, Faculty of Natural Sciences, University of Haifa,
Haifa 3498838, Israel}



\begin{abstract}

Line locking (LL) of absorption line systems is a clear signature of the dynamical importance of radiation pressure force in driving astrophysical flows, with recent findings suggesting that it may be common in quasars exhibiting multiple intrinsic narrow absorption-line (NAL) systems. In this work we probe the phase space conducive to LL and follow the detailed kinematics of those systems that may lock at the velocity separation of the \ion{C}{4}\,$\lambda\lambda 1548.19,1550.77$ doublet. We find that a small volume of the phase-phase admits LL, suggesting a high-degree of fine-tuning between the physical properties of locked  systems. The stability of LL against quasar luminosity variations is quantified with implications for the long-term variability amplitude of quasars and the velocity-separation statistic between multiple NAL systems. The high occurrence of LL by the CIV doublet implies that the hidden extreme-UV emission from quasars is unlikely to be significantly under-estimated by current models. Further, the ratio of the LL velocity to the outflow velocity may serve as a powerful constraint on the composition of the accelerating medium. We conclude that LL poses significant challenges to current theories for the formation of non-intervening NAL systems, and speculate that it may be a manifestation of expanding circumstellar shells around asymptotic giant branch (AGB) stars in the quasar-host bulge.  

\end{abstract}



\section{Introduction} \label{sec:intro}

Gaseous outflows are ubiquitous in quasars, and are manifested as blueshifted resonance-line absorption with respect to the quasars' restframe \citep[][see also \citealt{zak14,leu19} for the detection of quasar outflows in emission]{cre03,ves03,gan08}. These are commonly detected in the rest UV through X-ray energies, and span a velocity range of $10^3-10^5\,\mathrm{km\,s^{-1}}$ \citep{cre03,kri18,rev20}. The outflow phenomenon is intimately linked to the physics of quasars and the super-massive black holes that power them \citep{bre18}. Further, absorption-line phenomenology implies significant amounts of metal-rich material that is expelled from the compact quasar environs, and may reach galactic and intergalactic scales \citep{ara18}. As such, the study of quasar outflows has implications for galaxy formation \citep{fab12,fio17,ros18,che22}, and the properties of the circum-/inter-galactic medium \citep{gas13,kau17,bar18, liu18}. 

The phenomenology associated with quasar absorption line systems is vast. Some systems appear narrow with velocity dispersions $\lesssim 10^2\,\mathrm{km~s^{-1}}$, and may consist of several distinct kinematic components \citep{cul19,che19}. Other systems exhibit broad ($\sim 10^4\,\mathrm{km~s^{-1}}$) absorption line (BAL) profiles, which may be broken into several narrower kinematic components, but often have smoother appearances \citep{pao13}. For the latter type, the large velocity spread, the detection of partial-coverage effects (implying small sizes with respect to the background continuum emitting region), and the occasional time-variability of the troughs \citep{gib10}, imply an association with the inner quasar engine. In contrast, several distinct origins exist for narrow absorption-line (NAL) systems \citep{mis07,che18}: some are associated with material dispersed over cosmological scales, while others, particularly those with velocities $\lesssim 20,000\,\mathrm{km\,s^{-1}}$ with respect to the quasar rest-frame, are likely associated with the quasar and its host galaxy \citep{fol86,nes08}, as is indeed supported by time-variability \citep{nar04,wis04,lu18} and partial coverage effects \citep{cre03}. 

The physics of quasar NAL outflows is poorly understood. For some systems, especially in low-luminosity sources where the gas outflows with velocities $\lesssim 10^3\,\mathrm{km~s^{-1}}$, it has been suggested that the absorbers are cool condensations perhaps embedded in a hot and compact, thermally expanding wind \citep{che05}. Another explanation associates the outflowing gas with a more extended, dust-driven medium \citep{wil20}. For NAL systems observed at higher velocities, it has been suggested that a fast wind with a high kinetic energy can shock the ambient interstellar medium, and push clouds to their observed velocities \citep{fau12,wat17,zei20}. Alternative explanations for high-velocity multi-component absorption systems suggest a compact origin in an accretion disk, whose emission drives a wind by means of radiation pressure force, largely due to line and continuum absorption \citep{kas13,nom13,hig14,que20}. The latter scenario is supported by the phenomenon of line locking (LL), which is the focus of the present work. Some variants of the aforementioned scenarios include also the effect of magnetic fields that can assist to launch the gas, collimate it, and promote the survival of cool condensations against evaporation and hydrodynamic instabilities \cite[][in the context of broad-line flows]{dek95,eve05}. Understanding which of the above mechanisms is relevant to which type of NAL systems has significant implications for feedback and accretion-disk science \citep[][and references therein]{lah21}.

The paper is organized as follows:  in \S\ref{sec:ll} we outline the properties and physics of line-locked systems. The steady-state conditions conducive to line-locking are explored in \S\ref{sec:phase}. The kinematics of line-locked systems are further explored in \S \ref{sec:kin}, and further constraints on the available phase-space for line-locking are outlined. The discussion follows in \S \ref{sec:dis}, where the implications of our results for outflow models are provided. A summary is provided in \S\ref{sec:sum}. %Unless otherwise stated, we focus on the LL phenomenon of the \ion{C}{4}\,$\lambda 1548,1550$ doublet.

\begin{figure}[t!]
\epsscale{1.18}
\plotone{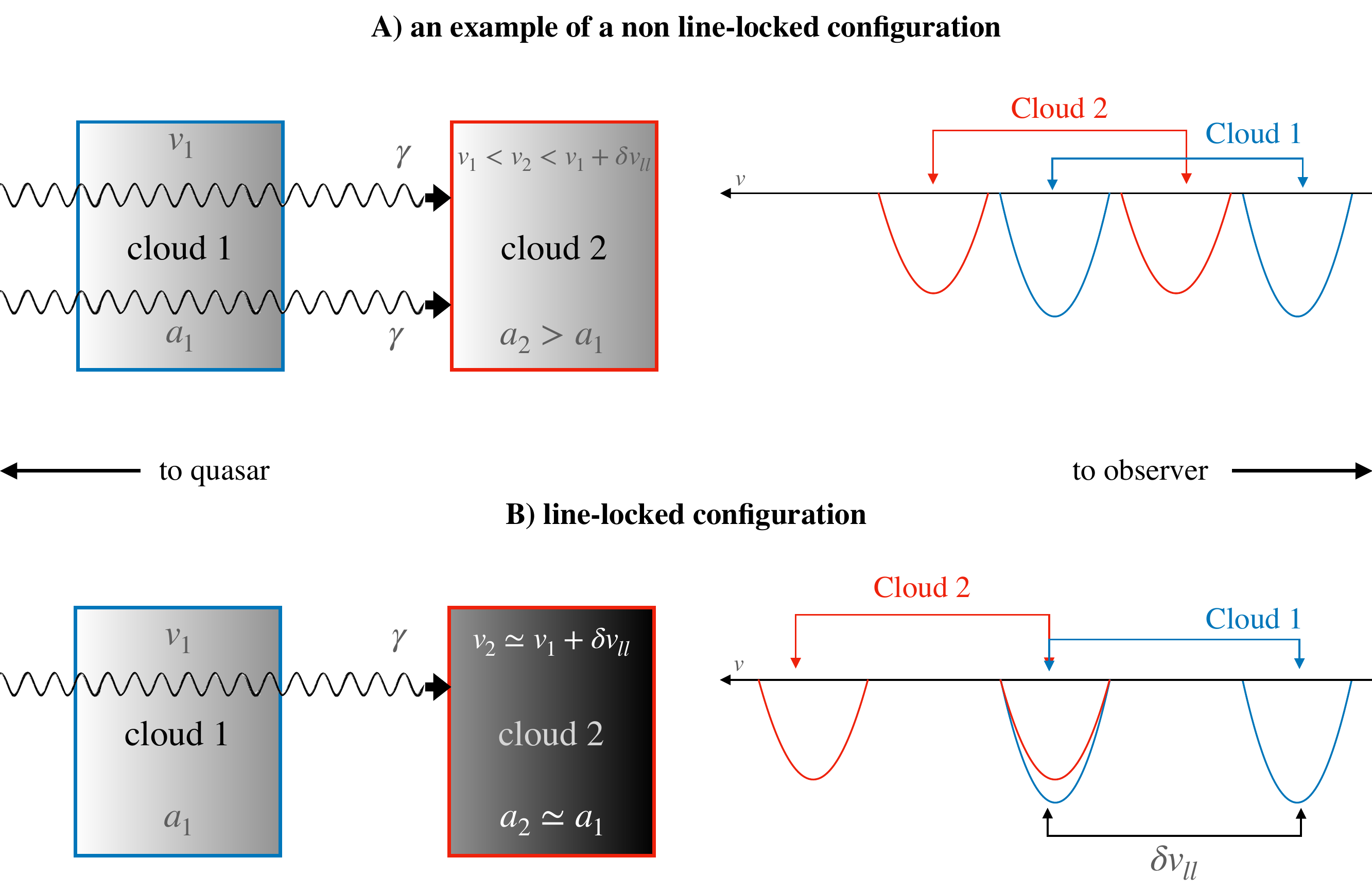}
\caption{A model for line-locking wherein two clouds are exposed to ionizing radiation from the left. The shielded cloud (cloud 2) has a higher acceleration than the shielding cloud (cloud 1), and is able to accelerate to higher velocities (compare the upper and lower panels) until the absorption-line troughs overlap in velocity space, its acceleration decreases to the point where the two clouds' accelerations are equal, and a line-locked position in acheived (lower drawing). \label{draw}}
\end{figure}

\section{Line-Locking}
\label{sec:ll}

Line-locking (LL) is a term describing a state in which the observed velocity difference between distinct kinematic absorption components along our sightline  equals the velocity-separation of known atomic transitions (Fig.~\ref{draw}). For example, \citet{ham11} reported multiple NAL systems toward a particular source, which are separated by the velocity difference of the \ion{C}{4}\ doublet \citep[see also][]{lin20a,lin20b,lu20}. LL is perhaps the clearest manifestation of the fact that radiative driving of gas is dynamically important in the astrophysical context \citep[][and below]{gol76}. 

\subsection{The phenomenology of LL NAL systems}

The absorption spectra of quasars are usually complex, with many NAL components present. Therefore, the detection of LL was historically limited to a small number of sources and its reliability and interpretation were, for many years, subject to much debate \citep{bor78,sar77,dre78,per78}. With the advance of large-scale high-resolution spectroscopy, many more LL systems were discovered \citep{tri97,sri00,sri02,sim10,gan13,che19}, and it was concluded that a velocity separation of $\simeq 500\,{\rm km~s^{-1}}$, which corresponds to the doublet separation of CIV\,$\lambda\lambda 1548.19,1550.77$, is relatively common \citep{sca70,bur75}. This was recently confirmed for the quasar population as a whole \citep{bow14,lu19,mas19,mas19b,che21}. Recently, statistical evidence for LL due to the \ion{Si}{4}\,$\lambda\lambda\,1393.76,1402.77$ doublet, at a separation of $\simeq 1900\,\mathrm{km~s^{-1}}$, was also reported \citep[][see also \citealt{fol87,sri02}]{lu19}.  LL which corresponds to the velocity difference between other transitions, such as \ion{O}{6}\ \citep{gan03,gan13}, \ion{N}{5}\ \citep{sri02,gan03,vei22}, and \ion{O}{6}-to-Ly$\beta$ \citep{gan13}, has been sporadically reported although it is not yet clear whether the small number statistics results from poor spectral resolution of large surveys, is due to chance coincidence in some studies, or results from a physical effect. 

\subsection{The physics of LL systems}

The conditions for LL were first described by \citet{mil26} in the context of radiation pressure acceleration of atoms in stars. \citet{mus72} coined the term ``line-locking", and applied it to radiatively accelerated gas in quasars \citep[see also][]{sca73}. A more complete treatment of LL was outlined by \citet{bra89}, which we now follow and extend. 

Consider two clouds that share the same sightline to a source of radiation and are accelerating away from it\footnote{LL can also occur between decelerating clouds; we do not consider such a scenario in the present work.}, with one of the clouds (cloud 1) shadowing the other (cloud 2). The shadow is wavelength dependent owing to the nature of absorption line cross-sections. If cloud 2 has some contribution to its total acceleration from a radiation pressure force term, $a_{\rm rad}(\lambda)$, which is wavelength (i.e., velocity) dependent, then line-locking will ensue provided 
\begin{equation}
    a\frac{\delta v_{ll}}{v}<a_2-a_1<\delta a_{\rm rad},
    \label{ll}
\end{equation}
where $a_1$ is the radiative acceleration on cloud 1, $a_2$ is the radiative acceleration on cloud 2, $v$ is the velocity of cloud 2 away from the source, and $\delta v_{ll}$ is the difference in the velocities of clouds 1 and 2 where line-locking occurs, which is equivalent to the difference in wavelength between the absorption features of CIV. The term $\delta a_{\rm rad}$ is the difference in the radiative acceleration between the state when cloud 2 is out of line-locked position (i.e., outside the shadow of absorption-line troughs due to cloud 1) to when it is maximally shadowed by it, i.e., it is aligned in velocity space with the center of the absorption-line trough due to cloud 1. In the latter position, line-driving is reduced due to shadowing (i.e., $\delta a_{\rm rad}>0$). The inequality on the right-hand side of Eq.~\ref{ll} means that the acceleration difference, $a_2-a_1$, flips sign between the shadowed and de-shadowed states so that the system could relax to an intermediate state, where $a_2=a_1$, and a fixed velocity difference between the clouds, $\delta v$ can be maintained so that $\delta v\simeq \delta v_{ll}$, where the latter term is the LL velocity, which is set by atomic physics. 

\begin{figure}[t!]
\epsscale{1.18}
\plotone{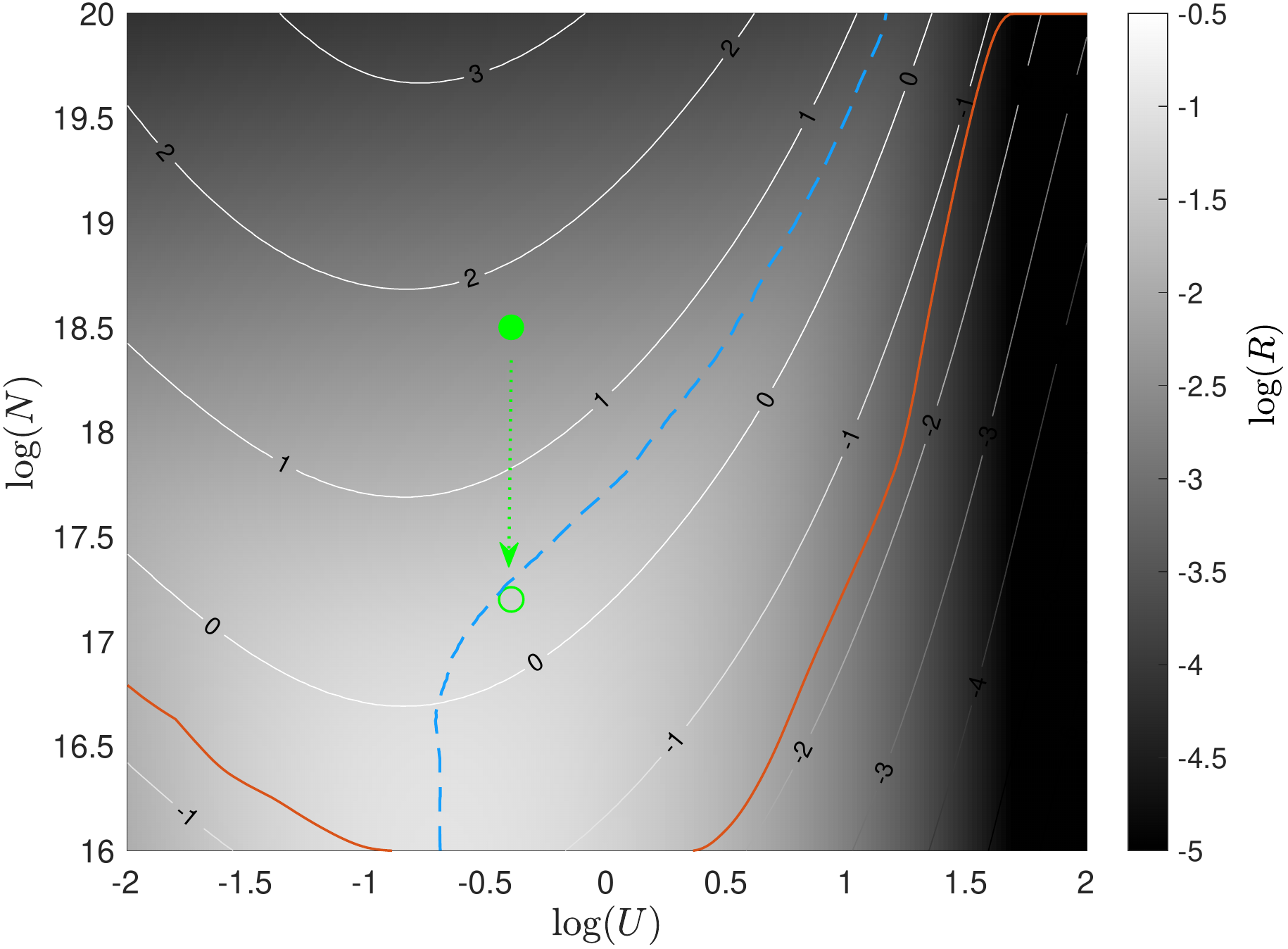}
\caption{The relative contribution of the \ion{C}{4}\,$\lambda\lambda 1548.19,1550.77$ doublet to the total radiation pressure force for a dusty medium. Brighter shades mark regions in phase space where the contribution is higher (see colorbar). Solid white lines mark iso-optical-depth contours with the optical depth at line center denoted in logarithmic units. The red solid (blue-dashed) curve marks the ridge in phase space over which $R$ is maximal when $U$ ($N$) is the independent variable. The solid green point marks the deduced model parameters from \citet{ham11}, while the empty green circle is a rough conversion of their results for the column per thermal width. \label{mom}}
\end{figure}

The left-hand condition in Eq.~\ref{ll} is set by the global kinematics of the outflow, whose outflow velocity $v$ is the average velocity of the two clouds, and is given to within a factor of order unity, by the ratio of the following dynamical timescales: the time it takes for cloud 2 to develop a relative velocity difference with respect to cloud 1, $\delta v_{ll}/(a_2-a_1)$, and the outflow dynamical time, $v/a$, where $a\equiv (a_1+a_2)/2$ is the average acceleration of the outflow, which is well defined for nearly co-spatial clouds of similar properties (see below) for which $(a_2-a_1)/a \ll 1$. For LL to be reached for clouds whose initial velocity difference is $\ll \delta v_{ll}$, it is required that $(\delta v_{ll}/v)/(a/(a_2-a_1))<1$. 

\begin{figure*}[ht!]
\epsscale{1.15}
\plottwo{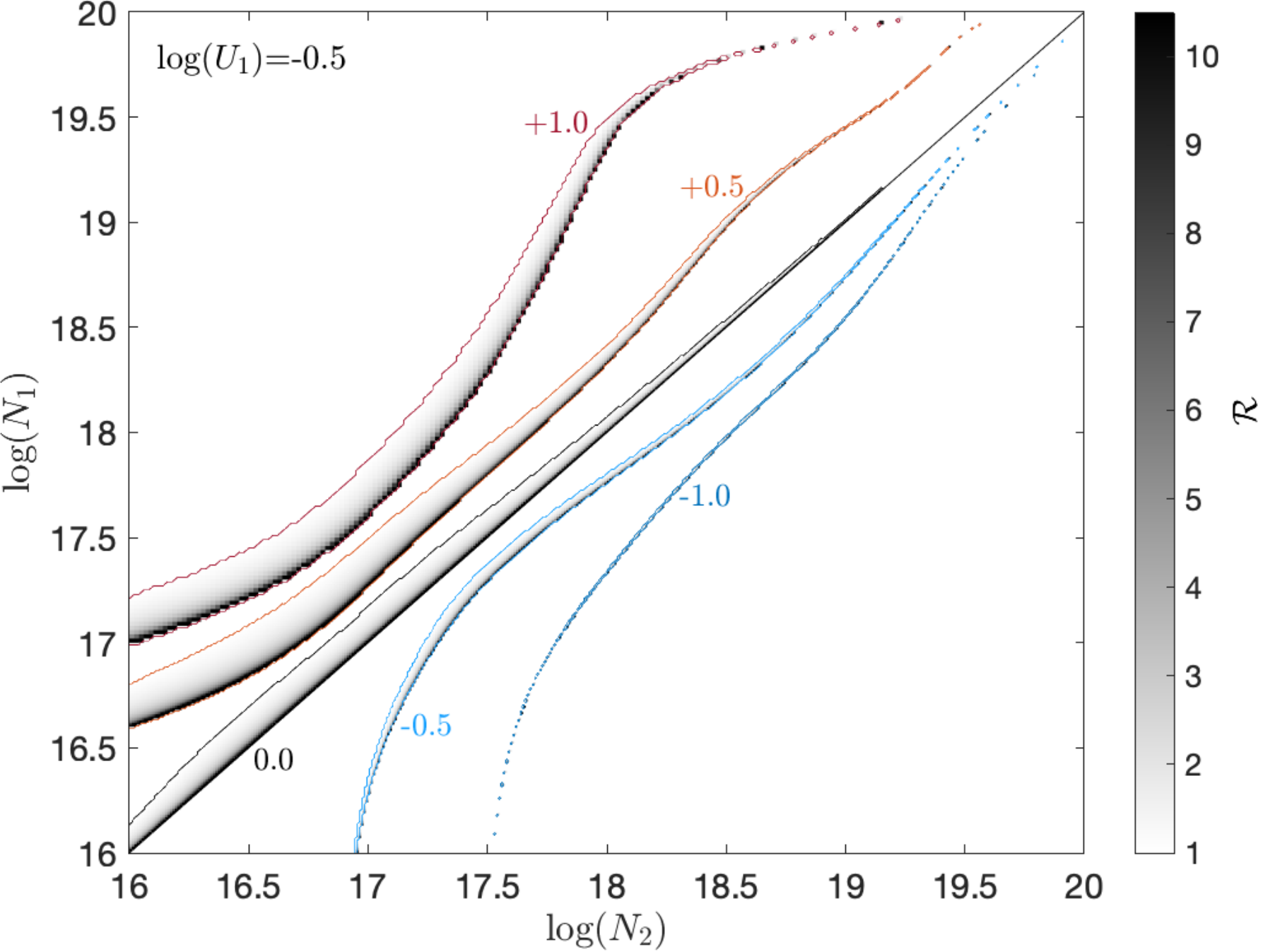}{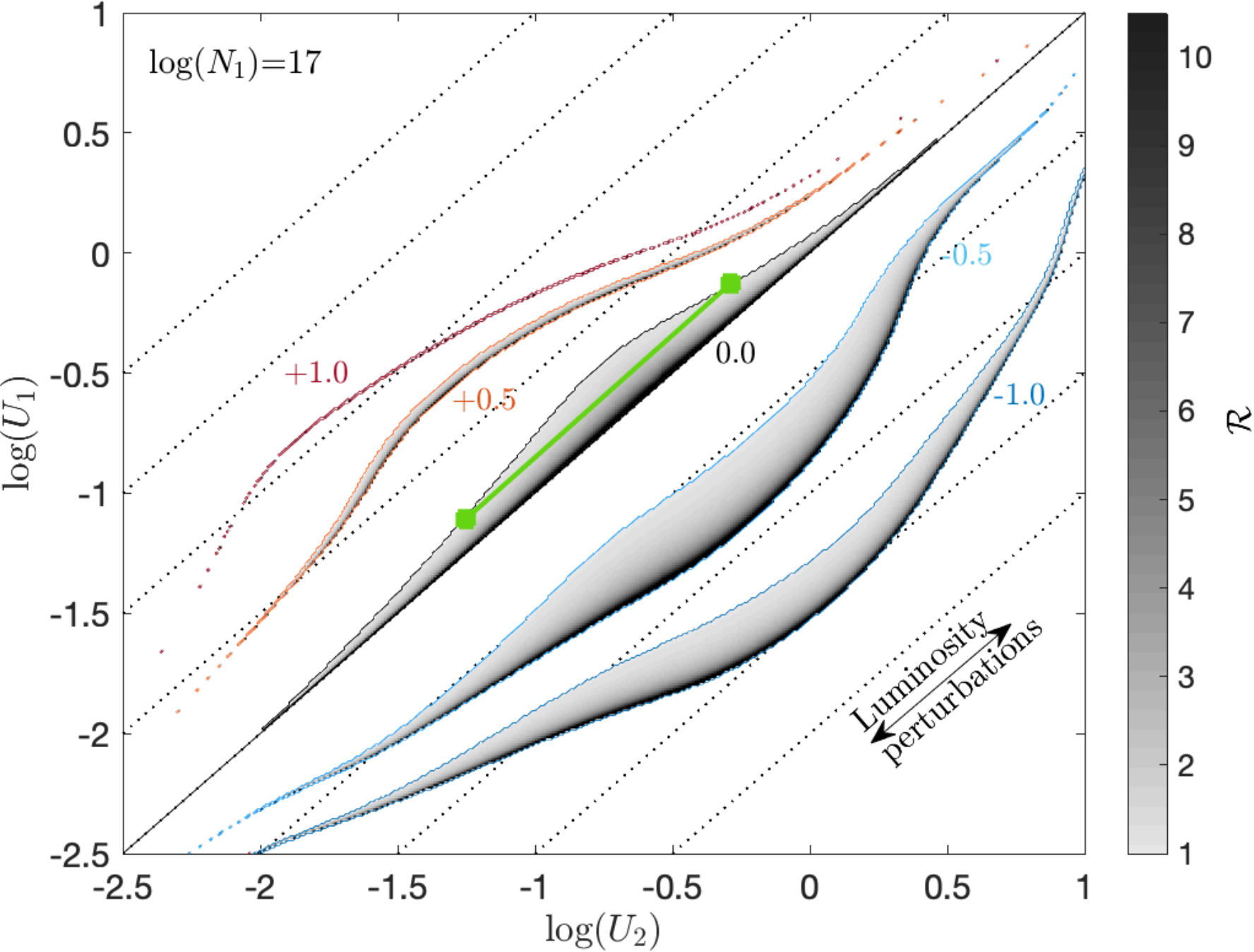}
\caption{The phase-space available for LL, as delineated by $\mathcal{R}$ with colored regions being characterized by $R\ge 1$ values (see the colorbars for color coding). {\it Left:} The allowed column-density phase-space under the assumption that the ionization parameter of the shielding cloud (cloud 1) is fixed. Several cases are depicted for different values of log($U_2/U_1$), which are denoted next to each colored surface. Discontinuities in the allowed phase space for a given log($U_2/U_1$)-value is the result of finite grid resolution. {\it Right:}  The allowed ionization-parameter phase-space (colored regions) under the assumption that the column-density of the shielding cloud is fixed, and for several values of log($N_2/N_1$), which are denoted next to each curve. Overlaid in dotted lines are trajectories along which luminosity perturbations occur. The green diagonal crossing the allowed phase-space for LL marks the median ionization parameter end-points, where LL models may be maintained (see \S\ref{qv}). \label{NNUU}}
\end{figure*}

\subsection{The properties of LL-NAL \ion{C}{4}\ systems}

Reliable constraints on the physics of LL systems are scant. In what follows we focus on the most common LL signatures in NAL systems that correspond to a velocity separation of $\delta v_{ll}\simeq 500\,\mathrm{km~s^{-1}}$ due to the CIV\,$\lambda\lambda 1548.19,1550.77$. For a particular system (J\,2123-005), \citet{ham11} concluded that the gas is highly ionized, with \ion{O}{6}\ being the abundant ionization state of oxygen, and implying an ionization parameter\footnote{The ionization parameter, $U$, is the ratio of the ionizing photon density to the electron density in the medium.}, $U\lesssim 1$ for a typical type-I quasar spectral energy distribution (SED), with a gas column density of $\sim 10^{19}\,{\rm cm^{-2}}$, and a slightly above solar metalicity with $Z\simeq 2Z_\odot$ ($Z_\odot$ is the solar composition). These authors also concluded that partial coverage effects are important and may be transition-dependent. This means that the clouds have sizes which are comparable to, or smaller than those that characterize the continuum emitting region in this source. For continuum emission originating from a standard accretion disk, the authors estimated absorber scales of $\sim 0.01$\,pc or smaller. The detection of variability in that system implied gas densities $>5000\,\mathrm{cm^{-3}}$ based on recombination-timescale arguments. An upper limit on the density of $\sim 10^8\,\mathrm{cm^{-3}}$ was deduced from the lack of discernible acceleration during the campaign \citep{ham11}. The above constraints imply a length scale for the absorbing material along our sightline of $10^{-7}-10^{-3}$\,pc, which when combined with partial coverage arguments, implies a spray of many small spherical cloudlets or a highly flattened sheet configuration for the outflow, with an aspect ratio of $10^{-5}-10^{-1}$. That such small structures exist in quasar outflows has been proposed in the context of BAL flows \citep[e.g.,][]{hal07}. The location of the outflowing material is rather poorly constrained but likely lies beyond the broad line region (BLR) and within the host galaxy's bulge. 

Large statistical samples \citep{bow14} demonstrated that there is dust associated with sightlines exhibiting LL systems, which leads to finite reddening at the level of $E(B-V)\simeq 0.005$\,mag per kinematic system. For dust typical of the interstellar medium (ISM), this corresponds to $A_V\sim 0.015$\,mag. If the gas composition of NAL absorbers is comparable to that of the galactic ISM, then such extinction levels correspond to column densities of $\gtrsim 10^{19}\,{\rm cm^{-2}}$ \citep{guv09}, which are not too different from the values reported by \citet{ham11}. These findings support an origin for the outflowing gas beyond the sublimation radius, in agreement with findings of \citet{ham11} based on an independent line of arguments.

The data presented by \cite{bow14} provide further clues into the structure of the region that leads to multi-component NAL absorption. After correcting for spurious signals and the contribution of intervening systems, these authors \citep[see also ][]{che21} find that the number of intrinsic NAL systems of a given multiplicity drops rapidly with the number of components found, such that the number of systems $\mathcal{N}_i$ of multiplicity $i$ (after averaging over the full velocity range), satisfies  $\mathcal{N}_1 \mathcal{N}_3/\mathcal{N}_2^2 \simeq 0.8\pm0.3$, which is consistent with the independent occurrence of clouds along the sightlines. Put differently, the coherence length of the medium that leads to NALs appears to be shorter than the physical separation between absorption components. Whether this applies also for LL-NAL systems is unclear although current statistics point to a $\gtrsim 50\%$ of multiple NALs, which are associated with the quasar being line-locked \citep{bow14}. 

\section{The Phase Space of LL Systems} \label{sec:phase}

The problem of the coupled dynamics of LL systems depends on the physical properties of each component, and therefore spans a multidimensional phase space. To simplify its treatment we first consider single cloud configurations, which do not admit LL.

\begin{figure*}[t!]
\plotone{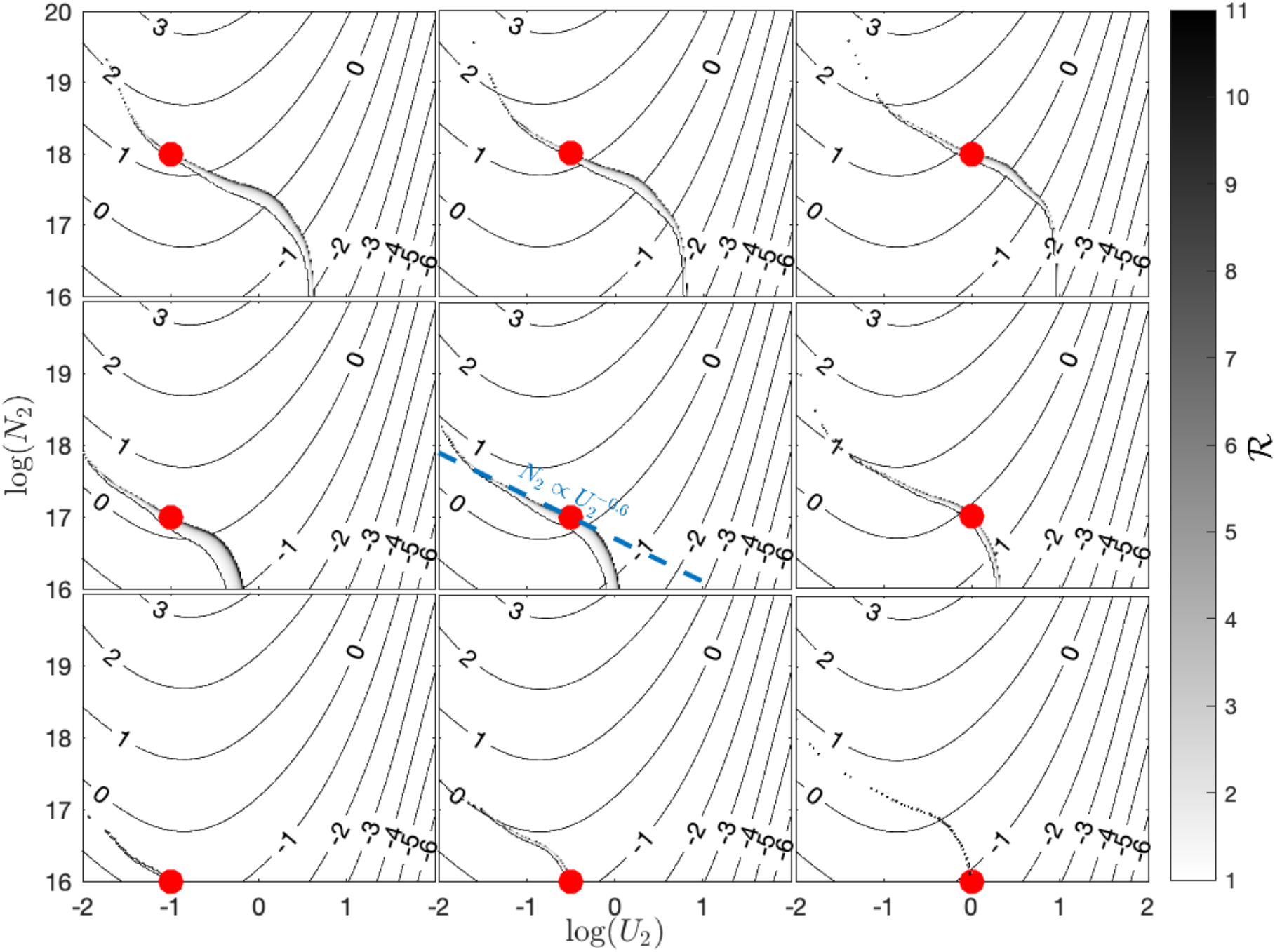}
\caption{The allowed LL phase-space for the shielded cloud (cloud 2), assuming particular properties for the shielding cloud (cloud 1), which are denoted by red points in the respective panels. Colored regions being characterized by $R\ge 1$ values (see the colorbar for color coding). Note the similarity in the shape delineating the relevant phase space for LL across all panels. Overlaid are optical-depth contours for the \ion{C}{4}\,$\lambda\lambda 1548$ transition from the shielded cloud. Also shown in dashed blue line is a slope, which roughly characterizes the shielded column density dependence on the ionization parameter for the allowed phase-space where the optical depth in the LL clouds is of order unity (shown only in the middle panel; see text). \label{ss}}
\end{figure*}

\subsection{Single-cloud Configurations}

Observations of LL systems imply that the \ion{C}{4}\,$\lambda\lambda 1548.19,1550.77$ contribution to the acceleration of the clouds is dynamically important, and it is clear from Eq.~\ref{ll} that larger values of $\delta a_\mathrm{rad}$ are more conducive to LL. Crudely, the observed visual extinctions \citep{bow14} imply that $\sim 1.5\%$ of the total radiative momentum carried by the quasar radiation field is deposited in the gas by dust, which is the dominant opacity agent. For saturated line-locked absorption, the total flux absorbed is $<\delta v_{ll}/c \sim 0.15\%$. Therefore $<10$\% of the total radiation pressure force may be deposited in the gas by the line-locked transition. More accurate estimations require detailed calculations that take into account the SED of the quasar continuum and all opacity and scattering agents of the accelerating gas, as we next outline.

Radiative acceleration is tightly linked to the composition of the accelerating medium, as well as to its ionization and thermal state, and its column. Throughout this work we assume isochoric clouds that are exposed to a  typical type-I quasar SED \citep{ham11}. The gas composition is set to twice the solar metal-to-gas value with ISM-like dust-to-metals ratio \citep[][ but see \citet{wu10} for higher values in NAL systems]{ham11,bow14}. Given the dilute nature of NALs with respect to typical critical densities of important transitions, the ionization and thermal state of the gas are fully determined by the ionization parameter, and the cloud's hydrogen column density, $N$. Given the low opacity associated with LL systems, isobaric cloud solutions should not lead to appreciably different results than those for isochoric ones. Likewise, models that include versions of radiation pressure confinement \citep{che01,bas14,ste14} will not exhibit significant compressions for optically thin highly-ionized dusty media as deviations of the radiation pressure force from the mean value are moderate across the cloud. 

The thermal and ionization state of the clouds is self-consistently calculated here using {\sc cloudy C17.01} photoionization code \citep{cloudy}. The total radiative acceleration was taken from {\sc cloudy}'s output and includes all major bound-bound, bound-free, free-free processes, as well as scattering by electrons and dust. The radiative acceleration by individual transitions was calculated ab initio assuming thermal broadening \citep{che03}. 

Figure \ref{mom} shows the ratio, $R$, of the radiation pressure force due the \ion{C}{4}\ doublet to the total radiation pressure force across the relevant phase space. We find that this ratio is maximized for $0.1\lesssim U\lesssim 1$, which is of order the observed values in J\,2123-005 \citep{ham11}. The observed column densities in this source are an order of magnitude larger than the calculated optimal columns for LL ($\sim 10^{17.5}\,\mathrm{cm^{-2}}$), but rough consistency is obtained when a correction is made for the suprathermal line broadening observed, which enhances the line-contribution to the radiative driving at larger columns (Fig.~\ref{mom}; see also \citealt{che01}). 

\subsection{Two-cloud Configurations}

Turning next to two-cloud configurations, which admit LL, we note the similar outflow velocity of LL systems that satisfy $\delta v_{ll}/v \ll 1$ ($\sim 0.05$ for the source studied by \citealt{ham11}), and the statistic of multiple NAL systems along our sightilne, which is consistent with their independent occurrence \citep[][and \S2.3]{bow14}. These motivate a model in which the two clouds are physically independent, and are characterized by distinct values for $U$, and $N$, but are approximately co-spatial, at least to the degree that differences in the geometric flux attenuation factors may be neglected. We further assume that the $a_2-a_1$ term in equation \ref{ll} is dominated by the difference in accelerations due radiation pressure force, and that the contribution of non-radiative terms, such as gravity and drag force, is negligible (but see \citealt{vil99} for the case of broad absorption line flows). To simplify the representation of the results, we take the limit $\delta v_{ll}/v \to 0$, which is relaxed later on. This implies that our findings for the phase-space volume available for LL may be over-estimated (see \S 4). Continuum shielding of cloud 2 by cloud 1 is neglected, which is justified for much of the phase space given the low-opacity of NALs. This approximation may be less accurate for some of the phase space, especially when large columns of low-ionization material are concerned, but should not affect the main conclusions presented in  this work. 

With the aforementioned setup and using the force-multiplier formalism \citep[][and references therein]{ara94,che01}, equation \ref{ll} takes the compact form 
\begin{equation}
    \mathcal{R}(U_1,N_1;U_2,N_2)\equiv \frac{\delta a_\mathrm{rad}}{a_2-a_1}=\frac{\delta M_\mathrm{CIV}}{\delta M}>1,
    \label{ll1}
\end{equation}
where the force multiplier, $M$, is the ratio of radiation pressure force due to all absorption and scattering processes to that due to electron scattering so that $a_\mathrm{rad}\equiv n_e \sigma_T L M/4\pi r^2 \rho c$. Here, $n_e (\rho)$ is the electron-number (gas-mass) density, $\sigma_T$ the Thomson cross-section, $L$ the bolometric luminosity of the quasar, $r$ the distance of the clouds from the ionizing source, and $c$ the speed of light. Therefore, $\delta M_\mathrm{CIV} \propto \delta a_\mathrm{rad}$, and $\delta M\propto a_2-a_1$. 

We calculate $\mathcal{R}$ over a wide range of ionization parameters and column densities, and map the regions in the four-dimensional phase space where Eq.~\ref{ll1} is satisfied. The ionization parameter and column density volume probed here is motivated by recent surveys suggesting that non-intervening NAL systems cover a $\gtrsim 3$\,dex range in the column density of prominent ions \citep{gan03,fech09,per16}, and a substantial, $>2$\,dex range in ionization-parameter values \citep{gan03,cul19}, which is echoed by theoretical calculations \citep{kur09,zei20}. We find that the phase-space volume, where LL can occur, is $\lesssim 1\%$ of the total phase-space volume considered here (logarithmic volumes are assumed throughout). Requiring that the optical depth in the \ion{C}{4} doublet exceeds unity so that its absorption signatures are clearly visible, reduces the fractional phase space volume to $\lesssim 0.5$\%. Further extending the phase space probed in terms of ionization parameters and/or column densities, and including additional constraints on the flow kinematics (e.g. on the ratio of LL velocity to the bulk velocity of the outflow, or robustness against quasar flux variations; see below) substantially reduces the relative phase space conducive to LL. Relaxing the assumption of co-spatiality of the clouds does not qualitatively change the above conclusion. At face value, this statistic contrasts the observed occurrence rate of line-locked systems among multiple NALs \citep{bow14}, and implies a physical process that greatly enhances LL. Below we quantify the requirements for LL to occur, and consider particular plane projections of the 4-dimensional phase space to map the implied correspondences between the LL clouds.

\subsubsection{The column-density plane}

We next consider two (cospatial) clouds that have identical ionization parameters, and hence densities. We choose $U=10^{-0.5}$, which is optimal for LL in our setup and consistent with the observations (Fig.~\ref{mom} and \citealp{ham11}). This scenario could arise, for example, in a thermally unstable medium, which bifurcates into cool and hot thermally stable phases \citep{mo96} with length-scales, hence columns, triggered by the perturbations' wavelengths. The phase space is shown in the left panel of Fig.~\ref{NNUU}. It is clear that for LL to be operating, the column densities of the clouds should be similar to $<50$\% under optimal conditions, with $N\lesssim 10^{17}\,\mathrm{cm^{-2}}$. Averaging over the allowed phase space, column densities need be similar to within 10\% to allow for LL. Qualitatively similar results are obtained for other values of $U$ (not shown). 

Allowing for $U_1 \ne U_2$, we find that the columns must be different for LL to work, so that $a_1\simeq a_2$, and yet the phase space over which LL operates is confined to a relatively narrow strip in phase space from which columns cannot deviate by more than $\simeq 50$\% (Fig.~\ref{NNUU}). This is especially true in cases where the shielded cloud (cloud 2) is less ionized than the shielding cloud (cloud 1). In that case, the column densities must be follow a strict relation with deviations of order per-cent or less for LL to operate.

\subsubsection{The ionization-parameter plane}

We now consider a scenario in which $N_1=N_2$ (right panel of Fig.~\ref{NNUU}). We find that $\mathcal{R}>1$ for $U_1\simeq U_2$. In particular, for column densities leading to an optical depth of unity in the \ion{C}{4}\,$\lambda\lambda 1548.19,1550.77$ lines, any density differences between the absorption systems must be at most $50$\% for LL to operate, and often much smaller than that. Assuming $N_1\ne N_2$, the allowed phase space defined in the $[U_1,U_2]$ plane considerably shrinks; cases in which $N_2>N_1$, $U_1$ and $U_2$ must be fine-tuned to within a few per-cents for LL to operate. Generally, the allowed phase-space is delineated by contours for which $U_2 \not\propto U_2$ with implications for LL stability over time (\S\ref{qv}).

\subsubsection{The ionization-parameter--column-density plane}

Figure \ref{ss} shows the phase-space available for LL when the properties of the shielding cloud, $[U_1,N_1]$, are fixed at several observationally motivated values for which the implied \ion{C}{4}-doublet optical depth is in the range 0.1-10. The phase space defined by the shielded cloud, $[U_2,N_2]$, for which Eq.~\ref{ll1} is satisfied, shows a similar (although not identical) behavior for the cases explored here. Specifically, the allowed phase space follows a ridge, which may be locally (crudely) approximated by a broken powerlaw form ($N_2(U_2)\propto U_2^{\eta}$) followed by an abrupt cutoff at high values of $U_2$. The cutoff results from the radiation pressure force decreasing with increasing ionization level so that even optically thin, low column-density gas cannot satisfy Eq.~\ref{ll1} beyond some value of $U_2$. The powerlaw index, $\eta \simeq -0.6$ for system properties similar to those found by \citet{ham11} with optical depths $\lesssim 10$, but becomes steeper ($\eta < -1.0$) for lower values of $U_2$ due to the rapid increase in the opacity of helium and hydrogen. As noted before, a high-level of fine tuning, of order a per-cent or less, is required between clouds when $N_2>N_1$. Overall, a lower degree of fine-tuning of the clouds properties is required when the optical depth in the \ion{C}{4}\,$\lambda\lambda 1548.19,1550.77$ doublet is of order unity.

\section{The Kinematics of LL Systems}
\label{sec:kin}

\begin{figure}[t!]
\epsscale{1.18}
\plotone{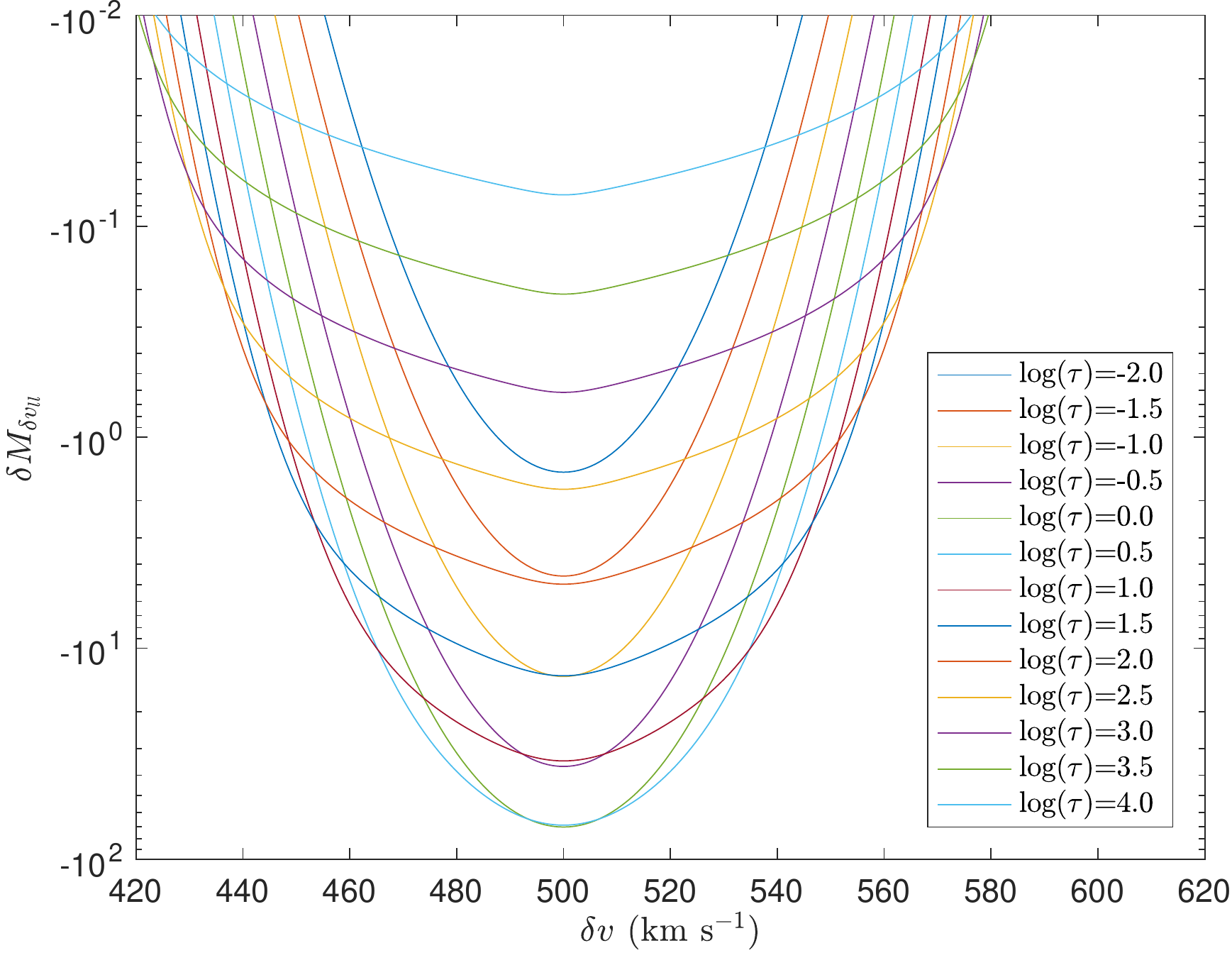}
\caption{The drop in the radiation pressure force (force multiplier) due to line blanketing, $\delta M_{\delta v_{ll}}$. Shown are calculations of Eq.~\ref{dMC4} between clouds with identical optical depths in the line troughs, and identical line-broadening with $\sigma=10\,\mathrm{km~s^{-1}}$. A bolometric correction of $\nu L_\nu/L=0.25$ was assumed. It is also assumed that $\tau_e=10^{-7}\tau$, where $\tau$ spans a range of values (see legend). The largest drop in radiation pressure force for each model is obtained when the systems overlap in velocity space at the doublet separation, $\delta v_{ll}\simeq 500\,\mathrm{km~s^{-1}}$. When comparing different models, clouds whose optical depths are of order unity, lead to the largest acceleration differences between fully-blanketed and non-blanketed configurations. \label{dM}}
\end{figure}

\begin{figure*}[t!]
\epsscale{1.18}
\plotone{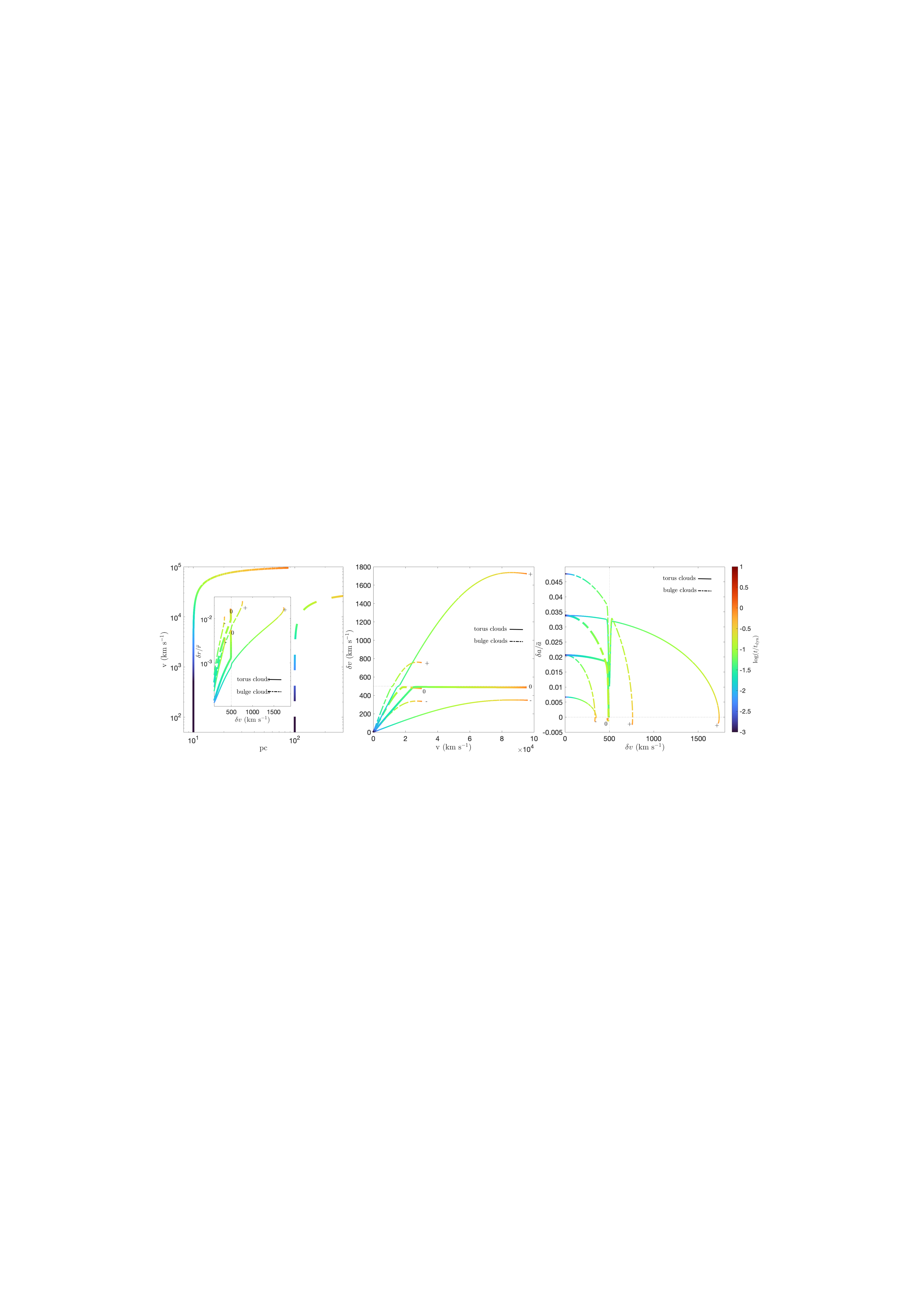}
\caption{Full kinematic solutions for clouds around LL conditions. Two sets of models are considered, which pertain to a high-luminosity quasar \citep{ham11}: clouds that are launched from dusty-torus scales (solid curves), and clouds that originate in the host's bulge (dashed curves). In all cases $N_1=N_2=10^{17}\,\mathrm{cm^{-2}}$ and $U_2=10^{-0.77}$. Three parameterizations are considered for the torus model: $U_1=10^{-0.61}$ (marked by ''$0$''), $U_1=10^{-0.72}$ (marked by ''$-$"), and $U_1=10^{-0.50}$ (marked by ''$+$'').  Three parameterizations are considered for the bulge model: $U_1=10^{-0.50}$ (marked by ''$0$''), $U_1=10^{-0.61}$ (marked by ''$-$''), and $U_1=10^{-0.39}$ (marked by ''$+$''). {\it Left:} velocity profiles with the inset showing the relative radial distance accumulated as a function of velocity difference. {\it Middle:} the velocity difference as a function of the outflow velocity. {\it Right:} the relative acceleration of the clouds as a function of the velocity difference. Note that small changes in the cloud properties (of order 30\% in $U_1$), which do not appreciably change the global clouds kinematics have a substantial effect on the ability to LL (see text). \label{lld}}
\end{figure*}

The above analysis is relevant for testing whether the observed properties of NALs can be maintained in a LL position under steady-state conditions, but do not reveal whether LL may be achieved in the first place, nor whether it may persist under time-varying conditions, such as near variable quasars. Here we treat the kinematic problem of two clouds by following their evolution from the launching point and until the coasting phase sets in. We focus on phase-space configurations that lead to LL. The coupled systems' kinematics follows from the equations of motion:
\begingroup
\renewcommand*{\arraystretch}{2.0}
\begin{equation}
    \begin{array}{ll}
        \displaystyle \dot{v_1}= & \displaystyle \frac{x_m}{m_p} \frac{\sigma_T L}{4\pi r_1^2 c}M_1(U_1,N_1) \\
        \displaystyle \dot{v_2}= & \displaystyle \frac{x_m}{m_p} \frac{\sigma_T L}{4\pi r_2^2 c}M_2(U_2,N_2; U_1,N_1, \delta v,\delta v_{ll})    \end{array},
    \label{eqmot}
\end{equation}
\endgroup \\
where $\delta v = v_2 - v_1$, $x_m=n_e/\rho \simeq 0.85$ for the assumed gas composition, and $m_p$ is the proton mass. In the numerical solutions presented below, this set of equations is solved for the kinematics of each of the coupled clouds. To assist with the interpretation of the results we note that the equation of motion for the velocity difference, $\delta v\equiv v_2-v_1$, between the clouds, in the limit of near co-spatiality ($\delta r/r \ll 1$ where $r=(r_1+r_2)/2$ and $\delta r=r_2-r_1$) is given by
\begin{equation}
\dot{\delta v}\simeq\frac{x_m}{m_p} \frac{\sigma_T LM}{4 \pi r^2} \left ( \frac{\delta M}{M} - \frac{2\delta r}{r} \right ),
\label{eqmot1}
\end{equation}
where we assumed that $M_1\simeq M_2 = M$ (i.e., the force multipliers characterizing the two clouds are similar to within a small correction).
The term $\delta M\equiv M_2-M_1$ consists of a sub-term, which does not depend on the clouds' relative velocity, $\delta M_0$, and a sub-term which is associated with line-blocking, $\delta M_{\delta v_{ll}}$ and is responsible for LL:
\begin{equation}
    \delta M=\delta M_0 (U_1,N_1,U_2,N_2)+\delta M_{\delta v_{ll}}(\delta v).
\end{equation}
If the $\delta M_0$-term includes the full contribution from all absorption and scattering processes, including all relevant absorption lines -- as would be the output of many photoionization codes -- then the term
{
\begin{equation}
     \delta M_{\delta v_{ll}}(\delta v)=-\frac{1}{\tau_{e,2}}\frac{\nu L_\nu}{L} \mathcal{W}_1 * \mathcal{W}_2,
     \label{dMC4}
\end{equation}
where `$*$' denotes convolution with respect to velocity\footnote{ $\mathcal{W}_1 * \mathcal{W}_2 = \int_{-\infty}^\infty dv \mathcal{W}_1(v) {W}_2(\delta v-\delta v_{ll}-v)$.} and
\begin{equation}
     \mathcal{W}_i(v)=\frac{1}{\sqrt{c}} \left [ 1-\mathrm{exp} \left ( \displaystyle -\tau_i e^{\displaystyle -v^2/2\sigma_i^2} \right )  \right ],
     \label{dMC4W}
\end{equation}
where $\sigma_i$ is the thermal broadening velocity of cloud $i$} (our photoionization calculations in \S3 show that $\sigma_1\simeq \sigma_2$ for the conditions most conducive to LL). In the above expression we assume for all practical purposes that $\sigma_i,dv_{ll}\ll c$, and that the optical depth for electron scattering from cloud 2, $\tau_{e,2} \ll 1$. The optical depth at the line center, $\tau_i$, is such that $\tau_1$ is the optical depth at line center for the \ion{C}{4}\,$\lambda 1548.19$ transition from cloud 1, and $\tau_2$ is the optical depth at line center for the \ion{C}{4}\,$\lambda 1550.77$ transition from cloud 2. A Gaussian dependence of the optical depth from the line-center was assumed, as is appropriate for metal NALs.

The dependence of $\delta M_{\delta v_{ll}}$ on the velocity separation between the clouds is shown in Fig.~\ref{dM} for the case of equal optical depths in the relevant transitions, $\sigma_1=\sigma_2\equiv \sigma=10\,\mathrm{km~s^{-1}}$, and for a fixed ratio between the optical depth at the line center and that for electron scattering. As discussed in \S\ref{sec:phase} and shown in Fig.~\ref{dM}, the largest effect of line-blocking on the radiation pressure force is attained for optical depths of order unity and when $\delta v=\delta v_{ll}$. Specifically, for optical depths $>100$, $\vert \delta M_{dv_{ll}}\vert \sim 1$, and a high degree of fine-tuning of the clouds properties is required to achieve LL since $M\gtrsim 10^3$ (for dusty media). LL could occur if there exists $0\le dv\le dv_{ll}$, where $\delta M$ flips sign. For much of the relevant phase space, this velocity lies in the range $450\le dv\le 500\,\mathrm{km~s^{-1}}$; the range is asymmetric with respect to $\delta v_{ll}$ since clouds that accelerate and develop an increasing velocity gap will lock first at $\delta v \le \delta v_{ll}$.\footnote{ Locking at $\delta v>\delta v_{ll}$ is not a stable equilibrium for differentially accelerating clouds but is a stable configuration for differentially decelerating ones.} In the limit of optically thin clouds of a fixed $\tau_e$ with $\sigma_1=\sigma_2$ and $\delta v=\delta v_{ll}$, equation \ref{dMC4} simplifies to a quadratic dependence on the optical depth such that $\delta M_{\delta v_{ll}}(\tau) \propto \tau_1\tau_2$, which is valid to within a factor of $\sim 2$ also for $\tau\simeq 1$. For $\tau \gg 1$, $\delta M_{\delta v_{ll}}$ has a square root logarithmic dependence on the optical depth, which is reminiscent of the curve-of-growth. 

\subsection{Kinematic Solutions Under Stationary Conditions}

It is beyond the scope of this work to span the full range of solutions for LL systems, and we focus on those solutions that appear to be more relevant to the observed systems. To this end we consider the emergence of LL in objects similar to J\,2123-005 \citep{ham11}, which are defined by a typical type-I quasar SED\footnote{{\sc Cloudy}'s AGN model was defined with the following parameterization: $T=10^5\,\mathrm{K},~\alpha_{ox}=-2,~\alpha_{uv}=-0.5$, and $\alpha_{x}=-1$.} with a bolometric correction of $\nu L_\nu(1500\,\text{\AA})/L=0.25$ and $\alpha_{ox}\simeq -1.9 $ \citep{ham11}, and with $L=8\times 10^{47}\,\mathrm{erg~s^{-1}}$. Gravity is neglected which is consistent with dusty media having $M\sim 2000-3000$ over the observationally relevant phase space, and with luminous quasars emitting close to their Eddington rate ($\Gamma_\mathrm{Edd}\simeq 1$), so that radiation pressure acceleration is highly effective even when the bulge mass is taken into account at large distances and so long as $M>200/\Gamma_\mathrm{Edd}$ \citep{kor13}, which may not be true for low-luminosity sources \citep{gav08}.

Cloud dynamics is treated ballistically. That is, the clouds are considered as distinct and coherent entities whose interaction with the environment -- e.g., with an ambient medium via drag forces -- is minimal, and does not lead to cloud disruption. Therefore, sonic/critical points in the solution are irrelevant. Further, we assume that the cloud properties ($U,N$) do not evolve with time. This assumption is not inherent to the model, but is employed for tractability of the problem given the multi-dimensional nature of the phase space. Lastly, special relativistic effects are ignored despite the high outflow velocities achieved by some models (e.g., Fig.~\ref{lld} for systems originating from torus-scales).

\begin{figure}[t!]
\epsscale{1.18}
\plotone{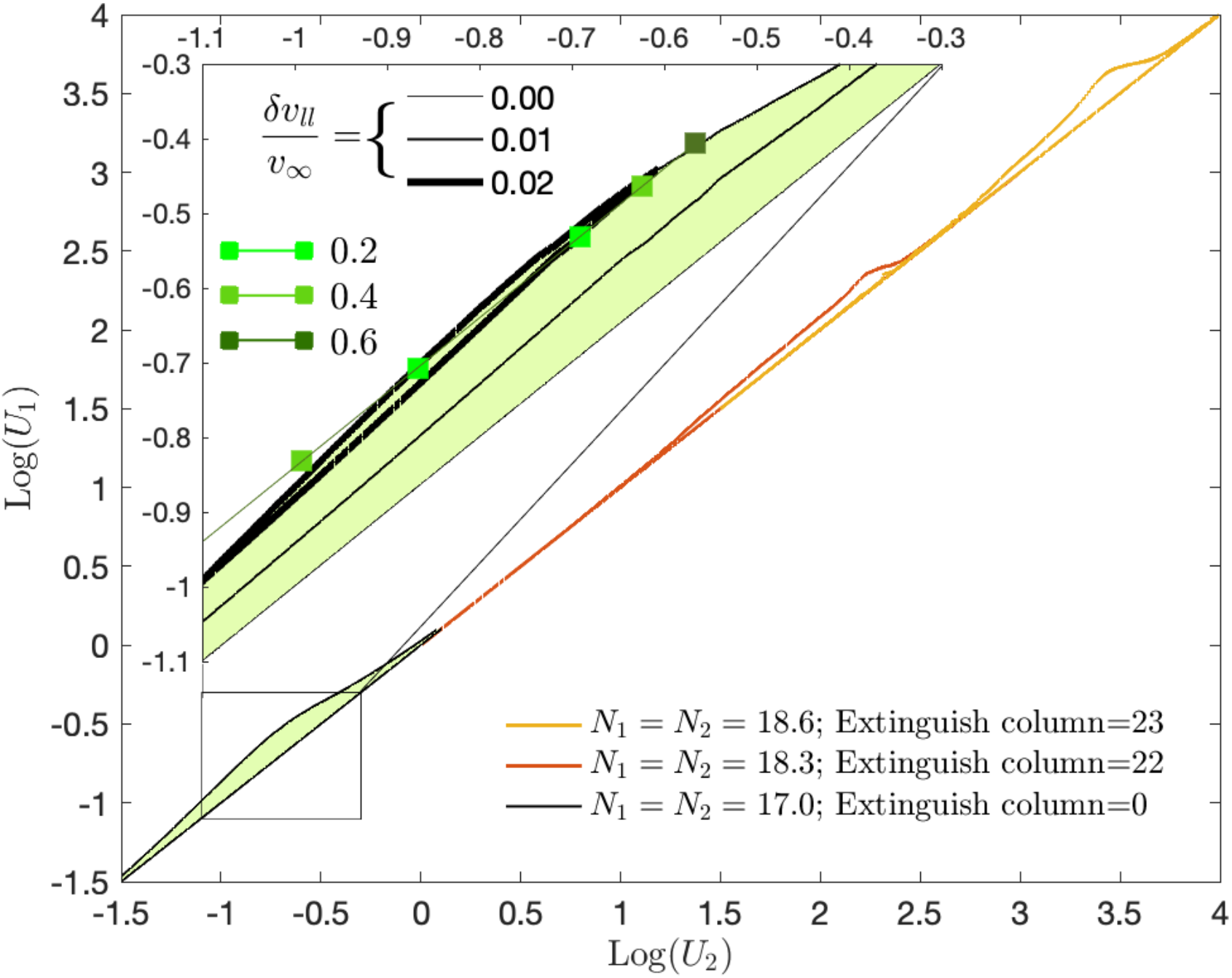}
\caption{The allowed LL phase-space for shielded clouds. The allowed LL phase-space is shown for clouds with equal column densities under several degrees of shielding (denoted by the shielding columns of neutral cold gas; see legend). Enhanced shielding pushing the optimal phase space for LL to higher ionization parameters (lower density gas), and higher columns (see text), but the area available for LL remains comparable. The inset shows a blow-up of the phase space for non-shielded clouds when a condition on the dynamical times is added (see text), for several levels of $\delta v_{ll}/v_\infty$. The particular clouds properties used in our kinematic analysis of \S\ref{qv} are denoted by colored stars. Quasar flux variations result in the clouds' properties tracing a diagonal line in the $[U_1,U_2]$ plane, whose length is defined by the RMS level of the sinusoidal signal (color coded; see inset's legend with values given in RMS over mean units). \label{NNUt}}
\end{figure}

We first consider a model in which the clouds are launched from the dusty region that lies just beyond the broad-line region -- the putative torus -- which we set to be at 10\,pc from the ionizing source \citep{bur13}. The model is characterized by $U_1=10^{-0.61},~U_2=10.^{-0.77}$ with $N_1=N_2=10^{17}\,\mathrm{cm^{-2}}$, and falls within the phase-space conducive to LL (Figs.~\ref{NNUU},~\ref{NNUt}). Calculations show that the clouds settle to a LL position within $<10\%$ of their dynamical timescale, and remain so out to their coasting phase. There is a subtle decrease in $\delta v$ with time owing to the growing radial distance, which results in $\delta r/r$ being comparable to $\delta M_{\delta v_{ll}}/M_2$, and resulting in a slight ''climb" of the shielded cloud along the absorption-line wing to reach a refined LL position. A model identical to the above but with $U_1=10^{-0.72}$ does not satisfy Eq.~\ref{ll} since the dynamical time is too short for $\delta v\simeq \delta v_{ll}$ to develop, and the clouds settle to $\delta v\lesssim 400\,\mathrm{km~s^{-1}}$ at their coasting phase. An identical model but with $U_1=10^{-0.5}$ does not settle to a LL position since the right-hand side of Eq.~\ref{ll} is not satisfied, and despite the decrease in acceleration seen at $\delta v\simeq 500\,\mathrm{km~s^{-1}}$, the clouds experience a monotonic relative acceleration to settle into a $\delta v\lesssim 2000\,\mathrm{km~s^{-1}}$ at their coasting phase (Fig.~\ref{lld}). 

We next consider a model in which the clouds are launched from the host galaxy's inner bulge at a distance of 100\,pc from the ionizing source. We first consider a model similar to the above but with $U_1=10^{-0.5}$, which formally does not admit LL (see above and Fig.~\ref{NNUt}). Nevertheless, LL still occurs for bulge clouds since cloud 2 develops a non-negligible radial gap with respect to cloud 1, so that $\delta r/r>\delta M/M$ (Eq.~\ref{eqmot1}). In particular, the system reaches a steady-state with $\delta v \lesssim \delta v_{ll}=500\,\mathrm{km~s^{-1}}$ after $\sim 30$\% of the dynamical time, whereupon the clouds accelerate nearly coherently. As the clouds move out, the radial gap between them increases up to $\sim 2\%$ of the distance to the ionizing source, thereby leading to a relative deceleration phase (right panel of Fig.~\ref{lld}), and to a decreasing $\delta v$ until a steady-state is reached with $\delta v\simeq 480\,\mathrm{km~s^{-1}}$. Any further radial gap increase has no effect on the gas kinematics. The same model but with $U_1=10^{-0.39}$ does not lead to LL as the effect of line-blocking is too small to balance the relative radiative acceleration ($\delta M>0$), and $\delta v\simeq 800\,\mathrm{km~s^{-1}}$ is reached at the coasting phase. An identical model with $U_1=10^{-0.61}$ fails to reach LL since the dynamical time to develop $\delta v_{ll}$ is longer than the outflow time in this case. To conclude, the phase-space diagrams shown in Fig.~\ref{NNUU},~\ref{ss} are indicative of the phase space volume conducive to LL, and yet the exact range depends on the launching site of the clouds via the left-hand side of Eq. \ref{ll}. 

\subsubsection{Kinematic constraints from LL systems}

LL introduces a further dynamical constraint, which can be used to recover some of the flow attributes, under the assumption that its observed properties are identical to those during the acceleration phase. Further assuming optically thin media, which is LL at velocity $\delta v_{ll}$, and has reached its terminal velocity, $v_\infty$, then the condition (Eq.~\ref{ll})
\begin{equation}
\frac{\delta v_{ll}}{v_\infty} < \frac{\delta M}{M}<\frac{\left \vert \delta M_{\delta v_{ll}}(\delta v= \delta v_{ll}) \right \vert }{M}, 
\label{ineq}
\end{equation}
for nearly co-spatial clouds translates to the following upper limit on the launching distance, $r_0=100r_\mathrm{100pc}$\,pc \citep[see  Eq.~\ref{vinf} below]{che01},
\begin{equation}
    r_\mathrm{100pc}<{ 10} L_{48} \frac{\tau_1^2 \tau_2^2}{M_3^{1/2}}\left ( \frac{f_\mathrm{CIV}}{{ 0.1}} \frac{Z_C}{{ 2}} \right )^2 \left ( \frac{b_L}{{ 4}} \right )^{-2} T_4 
\end{equation}
where $L_{48}\equiv L/(10^{48}\,\mathrm{erg~s^{-1}})$, the gas temperature, $T=10^4T_4$\,K, and $M=10^3M_3$. The factor $f_\mathrm{CIV}$ is the ionization fraction of \ion{C}{4}, and $Z_C$ is the abundance of carbon relative to the solar composition (of cloud 2). In the above expression $b_L$ is the bolometric correction with respect to the monochromatic UV luminosity ($b_L\simeq 4$ for the chosen SED). It was assumed that $\delta M_0<\vert \delta M_{dv_{ll}} \vert$ so that LL can be realized. For the particular case of J\,2123-005, $\tau_1\simeq \tau_2\simeq 1$, and assuming $f_\mathrm{CIV}=0.1$ and $Z_C=2$ \citep{ham11} and $M_3=2$ (as verified by photoionization calculations), we obtain $r_\mathrm{100pc}<5$ based on LL kinematics, which is consistent with the distance range reported by \citet{ham11} of 5-1100\,pc based on independent arguments. These kinematic arguments complement launching distance estimations based on the global outflow kinematics, where the asymptotic velocity satisfies \citep{che01},
\begin{equation}
    v_\infty\simeq 3\times 10^4 L_{48}^{1/2}M_3^{1/2}r_{100\mathrm{pc}}^{-1/2}\,\mathrm{km~s^{-1}},
    \label{vinf}
\end{equation}
and, conversely, the launching radius,
\begin{equation}
    r_{100\mathrm{pc}}\simeq 10 v_{\infty,4}^{-2} L_{48} M_3,
    \label{r100}
\end{equation}
where $v_\infty=10^4v_{\infty,4}\mathrm{km~s^{-1}}$. Here it was assumed that $M$ is constant along the acceleration path, which is reasonable given that radiation pressure acceleration by dust is less sensitive to the level of ionization of the gas and to its column density, so long as the medium is optically thin for dust absorption, which is justified for the column-density range observed. For the  case of J\,2123-005, LL estimates imply smaller scales by a factor of $\gtrsim 3$ than implied by global outflow kinematics ($r_{100\mathrm{pc}}\simeq 14$), suggesting that the outflow and/or quasar properties may be different over dynamical times than those implied by current observations.

\begin{figure}[t!]
\epsscale{1.18}
\plotone{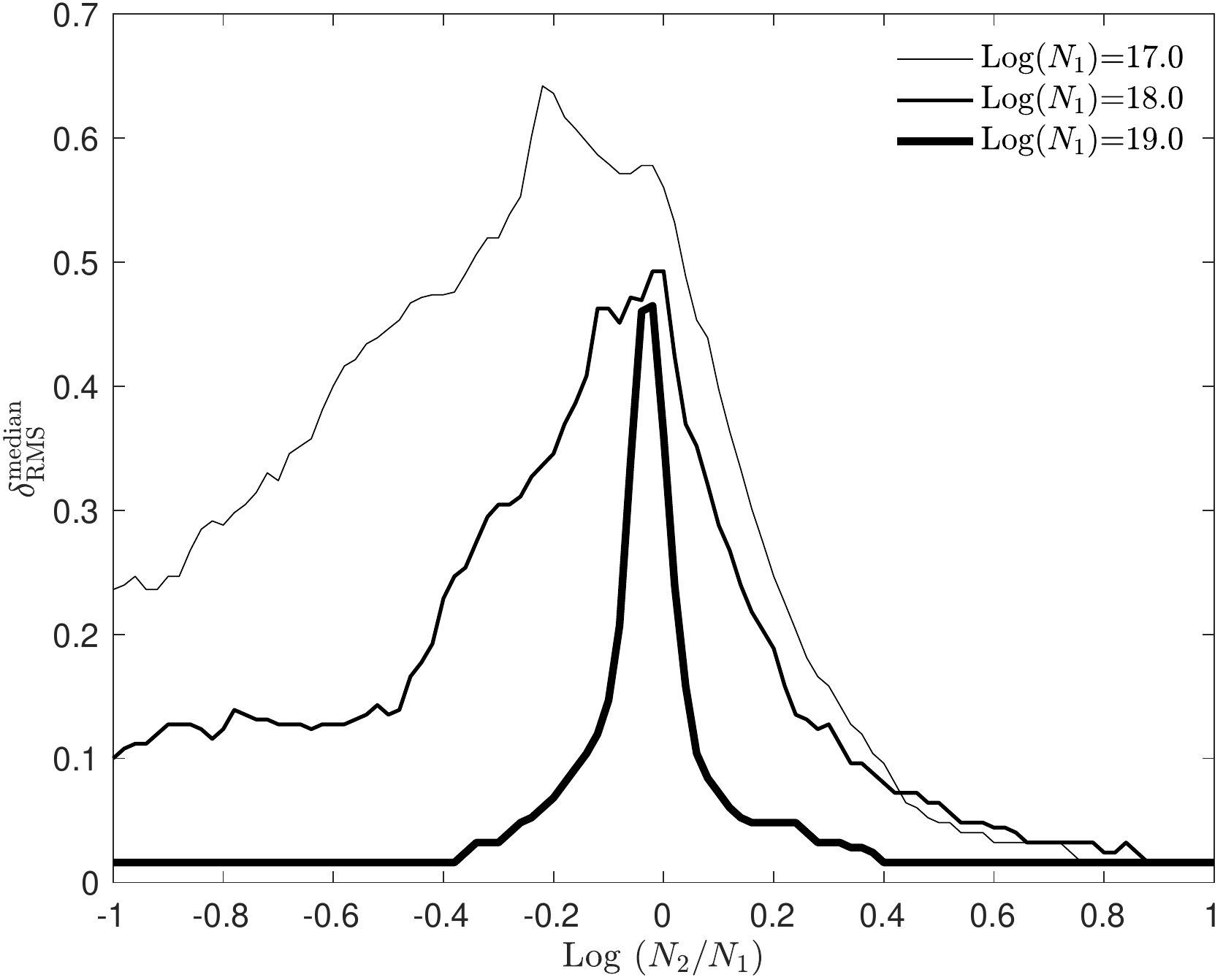}
\caption{Typical (median) peak-to-peak luminosity variations that are consistent with stable LL configurations, for three values of the column density of the shielding cloud (see legend in log units) and for a range of column densities of the shielded cloud (abscissa). For example, for cloud configurations with log$(N_\mathrm{H,shielding})=18$, log$(N_\mathrm{H,shielded})=17.5$, the median (over the available phase space, as appears in the right panel of Fig.~\ref{NNUU}) peak-to-peak that can sustain stable LL is $\simeq 0.2$.  \label{dLL}}
\end{figure}

\begin{figure*}[t!]
\epsscale{1.15}
\plottwo{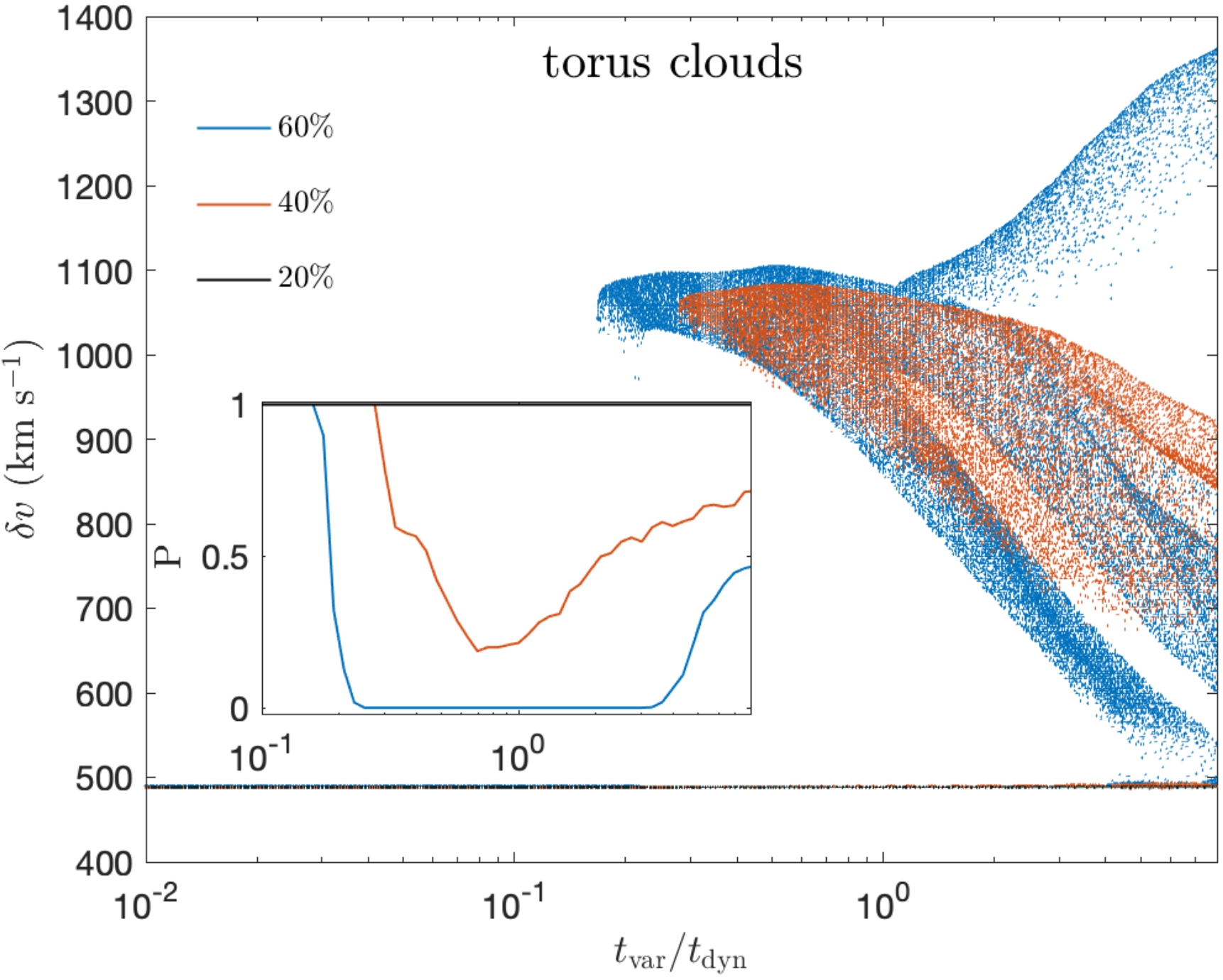}{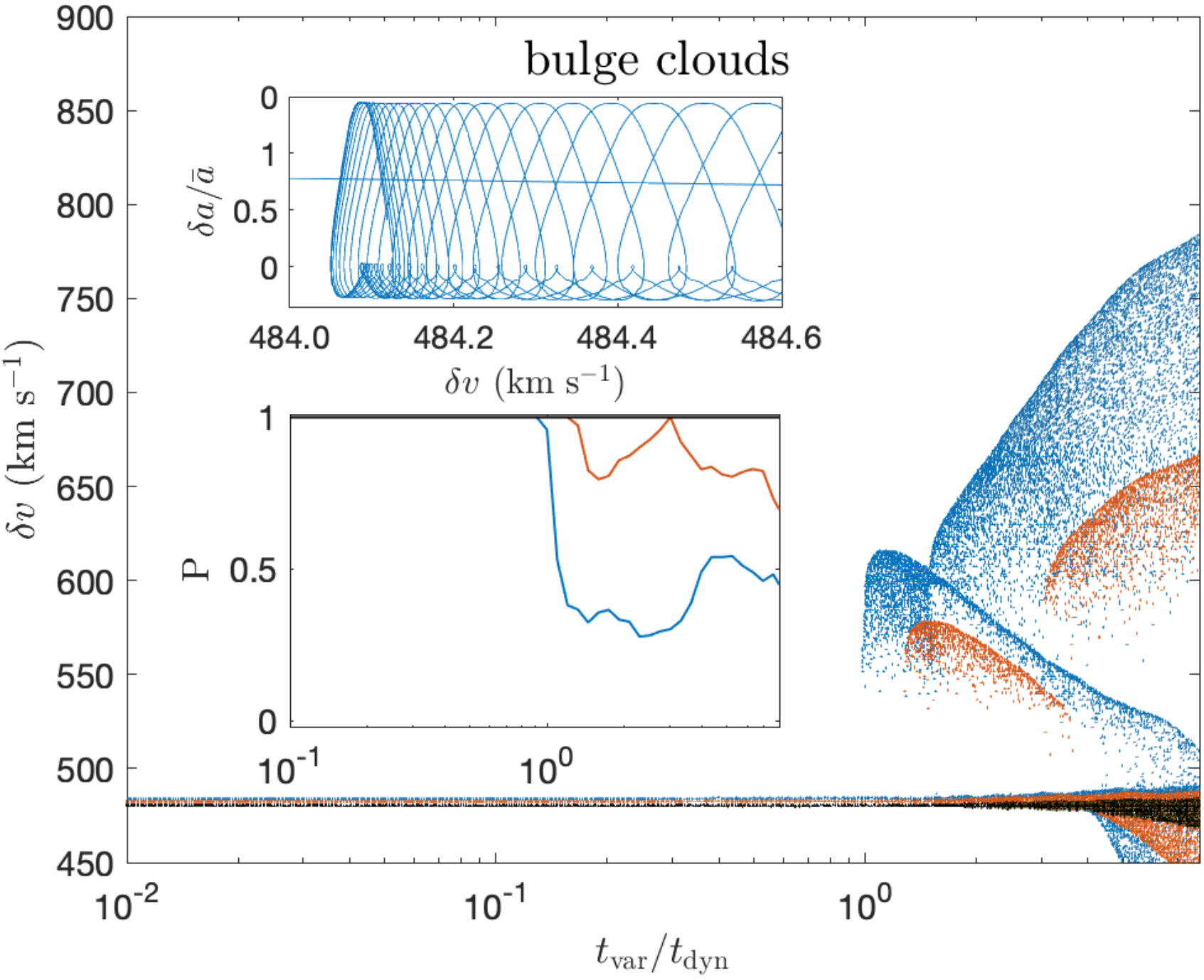}
\caption{Response of LL systems to quasar light variations (systems originating from the torus/bulge are shown in the left/right panel.  The asymptotic value of $\delta v$ between the systems are shown as function of the quasar variability period (assumed sinusoidal), $t_\mathrm{var}$ for several values of RMS variability (see legend in the left panel which applies to both panels). Systems can reach a LL position and maintain it when $t_\mathrm{var}$ is much shorter than all dynamical timescales. At periods comparable to or larger than dynamical timescales, the clouds do not settle to a LL position, and the the asymptotic $\delta v$ distribution shows a bifurcation pattern. The probability, $P$, for finding LL systems under quasar light variations of a given RMS amplitude with a period $t_\mathrm{var}$ is shown in the lower insets of both panels (here we defined a LL systems to have $450<\delta v<500\,\mathrm{km~s^{-1}}$). The upper inset in the right panel shows a typical solution for the response of LL clouds to periodic quasar variations (time flows along the blue curve with the clouds settling to a state with $\delta v\simeq 484.1\,\mathrm{km~s^{-1}}$, and show small velocity and relative-acceleration oscillations corresponding to an inverse ``heart" shaped curve). \label{qvar}}
\end{figure*}

\subsubsection{The role of continuum shielding}

It has been shown that substantial continuum shielding of quasar outflows can have significant dynamical effects \citep{mur95,che03}. Below we test whether extinguishing the continuum by a large column of neutral gas, which is external to the LL systems and lies along their sightline to the ionizing source, has a qualitative effect on the phase space available for LL. 

From Eq.~\ref{dMC4} it is clear that given optical depths in the lines, $\delta M_{\delta v_{ll}}$ inversely depends on the bolometric correction, which decreases when substantial (but Compton-thin) shielding columns are present. Still, revised bolometric factors are not expected to increase $\delta M_{\delta v_{ll}}$ by more than a factor of $\sim 2$. For the specific shielding scenarios simulated here, and an SED that peaks below the Lyman edge, changes to the bolometric correction are minor ($\sim 10$\%). The ratio $\tau_2/\tau_{e,2}$ also varies for substantial shielding columns since the relative fraction of ions at their maximum is typically lower and a wider range of ionization levels characterizes the ionized gas \citep{che03}. Our calculations show that the latter effect is dominant, and that shielding by large columns decreases the peak ratio of the \ion{C}{4}\ doublet radiation pressure force to the total radiation pressure force  to $\simeq 2$\% (compared to $\lesssim 10$\% for the non-shielded case; see Fig.~\ref{mom}). In comparison, changes to the total radiation pressure force are at the $\lesssim 10\%$ level between the shielded and non-shielded scenarios due to the dominance of dust opacity and the SED chosen, with $M \gtrsim 10^3$ in both cases.

The effect of shielding on the phase space available for LL is studied in Fig.~\ref{NNUt}, where the optical depth in the doublet lines is of order unity for preset values of $N_1=N_2$. Higher levels of shielding push the optimal phase-space range for LL to higher values of $U_1,~U_2$ (or, conversely, to lower densities), by as much as four orders of magnitude. However, the phase-space available for LL remains comparable in volume. Therefore, the effect of shielding does not alleviate the need for fine-tuning of the clouds properties to facilitate LL.

\subsection{Quasar variability and LL} \label{qv}

Quasars vary on a wide range of timescales, and are characterized by a red power spectrum, such that $P(\omega) \propto \omega^{-\alpha}$ with $2<\alpha<3$ over hours to years timescales \citep{dvr05,smi18}. Therefore, much of the variance is at the lowest frequencies with previous works suggesting substantial power on timescales of order $\sim 10^4$\,years \citep{kee17}, which are comparable to the outflow timescale:
\begin{equation}
    t_\mathrm{dyn}\equiv \frac{r_0}{v_\infty}\simeq 3\times 10^3\,L_{48}^{-1/2}r_{100\mathrm{pc}}^{1/2}M_3^{-1/2}\mathrm{years},
\end{equation}
where the force multiplier, $M=10^3M_3$. Another relevant timescale is the de-shadowing timescale over which the shielded cloud can accelerate relative to the shielding cloud by more than one thermal width, $\sigma/\delta a_\mathrm{rad}$, which is shorter than $t_\mathrm{dyn}$ by a factor of $\sigma/v_\infty$. 

The effect of quasar variability is to move a system of clouds defined in the $[U_1,~U_2]$ plane along $45$\textdegree\ diagonals (Fig.~\ref{NNUU}). Therefore, a model which satisfies the conditions for LL under steady-state conditions, may not do so if pushed by the fluctuating quasar flux to a region of phase-space that is not conducive to LL. Qualitatively, the larger the variability amplitude is, the more likely the system will be pushed away from LL equilibrium. 

To estimate the level of flux variations that may occur without disrupting LL, we consider three models for which $\mathrm{log}(N_1)=17,18,19$. For each of the models, we restrict the discussion to the range $0.1N_1\le N_2\le 10N_1$, and analyze the phase-space in the $[U_1,~U_2]$ plane in the following manner: for each set of $N_1,~N_2$ values, the phase space conducive to LL is calculated, which results in a simply connected surface in the $[U_1,U_2]$ plane. Each surface may be transected by $45$\textdegree\ diagonals of varying length, which is a measure of the peak-to-peak flux variation amplitude that maintains LL. The median length of all the transects is logged (see, for example, the green line in Fig.~\ref{NNUU} for a particular set of models, whose length corresponds to $\simeq 1$\,dex), and used as a measure for flux variability that may be tolerated by a pre-existing LL system. We note, however, that alternative measures may be defined, although these are less  probable to materialize under particular conditions are met (e.g., for systems that are nearly identical and lie along the diagonal). The above process is repeated for a range of $N_2/N_1$ values, and for each of the $N_1$ models defined above. We quantify the results by defining the median root mean square variability measure, $\delta_\mathrm{RMS}^\mathrm{median}$, where we assume that the quasar luminosity variations are of a sinusoidal form so that 
\begin{equation}
    L(t)=L_0[1+\delta \mathrm{sin}(\omega t+\phi)],
    \label{sin}
\end{equation}
where $\delta<1$ is the variation amplitude, and $\delta_\mathrm{RMS}\simeq 0.7\delta$. The angular velocity, $\omega\equiv 2\pi/t_\mathrm{var}$ where $t_\mathrm{var}$ is the period, and $\phi$ is a random phase (see below). 

The results are shown in Fig.~\ref{dLL} as a function of $N_1/N_2$ for several values of $N_1$. It is clear that low column density configurations have a higher tolerance to flux variations of the source, and among those, configurations for which the clouds columns are comparable (and hence their ionization parameters as well, as they lie along the main diagonal) are most robust. For larger columns, the system is relatively easy to disrupt from a LL equilibrium. Further, clouds configurations in which the shielding cloud has a higher column than the shielded cloud (i.e., $N_1>N_2$) are more resilient to luminosity fluctuations of the ionizing source. For the particular models shown in Fig.~\ref{dLL}, $\sim 30$\% flux variations are not expected to disrupt a pair of LL clouds with columns of order $10^{17}\,\mathrm{cm^{-2}}$ (assuming $\delta v_{ll}/v_\infty \to 0$), but could easily disrupt a system with whose columns are of order $10^{19}\,\mathrm{cm^{-2}}$ unless the columns agree to better than $\sim$10\%. We emphasize that the quoted results are likely upper-limits on the true susceptibility of the system to disruption since $\delta v_{ll}/v_\infty$ is finite, and the phase space conducive to LL is more limited; see Fig. \ref{NNUt} where larger $\delta v_{ll}/v_\infty$-values substantially reduces the available phase space and $\delta_\mathrm{RMS}$ (not shown). 

\subsubsection{Asymptotic inter-cloud kinematics}

As describe above, in our simulations we assume a single sinusoidal mode of a given amplitude, frequency, and random phase. Motivated by the data for J\,2123-005 \citep{ham11}, we consider two plausible models for the kinematics of dusty clouds: a model in which the clouds are accelerated from torus scales (10\,pc), and a model where they accelerate from bulge scales (100\,pc). The terminal outflow velocity for clouds traveling balistically with constant properties is that given by Eq. \ref{vinf}.

The inter-cloud kinematics follows from the solution to equations \ref{eqmot} with $L(t)$ given by Eq.~\ref{sin}. We assume the ionization-recombination timescales are the shortest in the problem so the gas thermal and ionization states are instantaneously set by $L(t)$, which translate to time-variation in $U$. We assume $N_1=N_2=10^{17}\,\mathrm{cm^{-2}}$, and $U_1=10^{-0.5},~U_2=10^{-0.77}$ (\S4.1). We track the velocity difference between the clouds, $\delta v$ at all times, and log its asymptotic value as a function of $\omega$ and $\delta$ for torus clouds and for bulge clouds.

As expected, luminosity variations on timescales much shorter than dynamical timescales ($t_\mathrm{var}\lll t_\mathrm{dyn}$) do not prevent clouds from attaining a LL position for $\delta_\mathrm{RMS}=0.7\delta\le 0.6$. In particular, the clouds relative acceleration, $\delta a$, and $\delta v$ traces closed loops in phase space while accelerating coherently to high velocities from bulge scales (Fig.~\ref{qvar}). Nevertheless, for luminosity variations that operate on timescales $t_\mathrm{var}\gtrsim 0.1t_\mathrm{dyn}$, and for $\delta_\mathrm{RMS}>0.2$, clouds do not settle, in most-to-all cases to a LL position, with their relative velocity showing a bifuraction pattern. The $\delta v$-range increases with $\delta_\mathrm{RMS}$. Qualitatively similar behavior is observed for torus clouds and for bulge clouds, although the $\delta v$-range in the latter is smaller on account of the smaller accelerations at large distances. For bulge clouds, $\delta v<\delta v_{ll}$ is also observed on account of the smaller relative clouds acceleration in a fraction of the models, preventing them from reaching LL velocities. In the latter models, the system is more stable to luminosity variations occurring on $t_\mathrm{var}\lesssim t_\mathrm{dyn}$, which is due to the fact that the system does not attain a line-locked configuration much before $t_\mathrm{dyn}$ (see above). 

The insets of figure \ref{qvar} also show the probability of a system of clouds to achieve a LL configuration under the effect of varying quasar luminosity for torus and for bulge clouds, as a function of the variability timescale. Clearly, the systems are most susceptible to variations over dynamical timescales. Systems that achieve their LL state over shorter timescales with respect to dynamical timescales are more prone to be driven out of LL equilibrium. In particular, for significant $\gtrsim 30\%$ variations over dynamical timescales, torus clouds with $t_\mathrm{dyn}\sim 10^3$\,years will all be driven out of LL equilibrium.

\subsection{Why LL of the \ion{C}{4}\ doublet?}

The \citet{bow14} and \citet{mas19} studies show that LL at the velocity separation of the \ion{C}{4}\,$\lambda\lambda 1548.19,1550.77$ doublet is common among multi-component intrinsic NALs, but find little evidence for substantial LL features at $\delta v_{ll}<500\,\mathrm{km~s^{-1}}$. While this could be partly attributed to the limited spectral resolution of large spectroscopic surveys, we are not aware of a large number of such cases found in high-resolution data. Conversely, LL with $\delta v_{ll}>500\,\mathrm{km~s^{-1}}$ have been sporadically detected due to transitions in the near UV \citep[e.g.,][]{sri00,lu18}, but do not appear to be as common as the CIV doublet locking \citep{bow14}. LL with velocity separations corresponding to far-UV (FUV) transitions, which lie at the presumed peak of of quasar emission, have not been, thus far, robustly identified to the best of our knowledge.

\begin{figure}[t!]
\epsscale{1.18}
\plotone{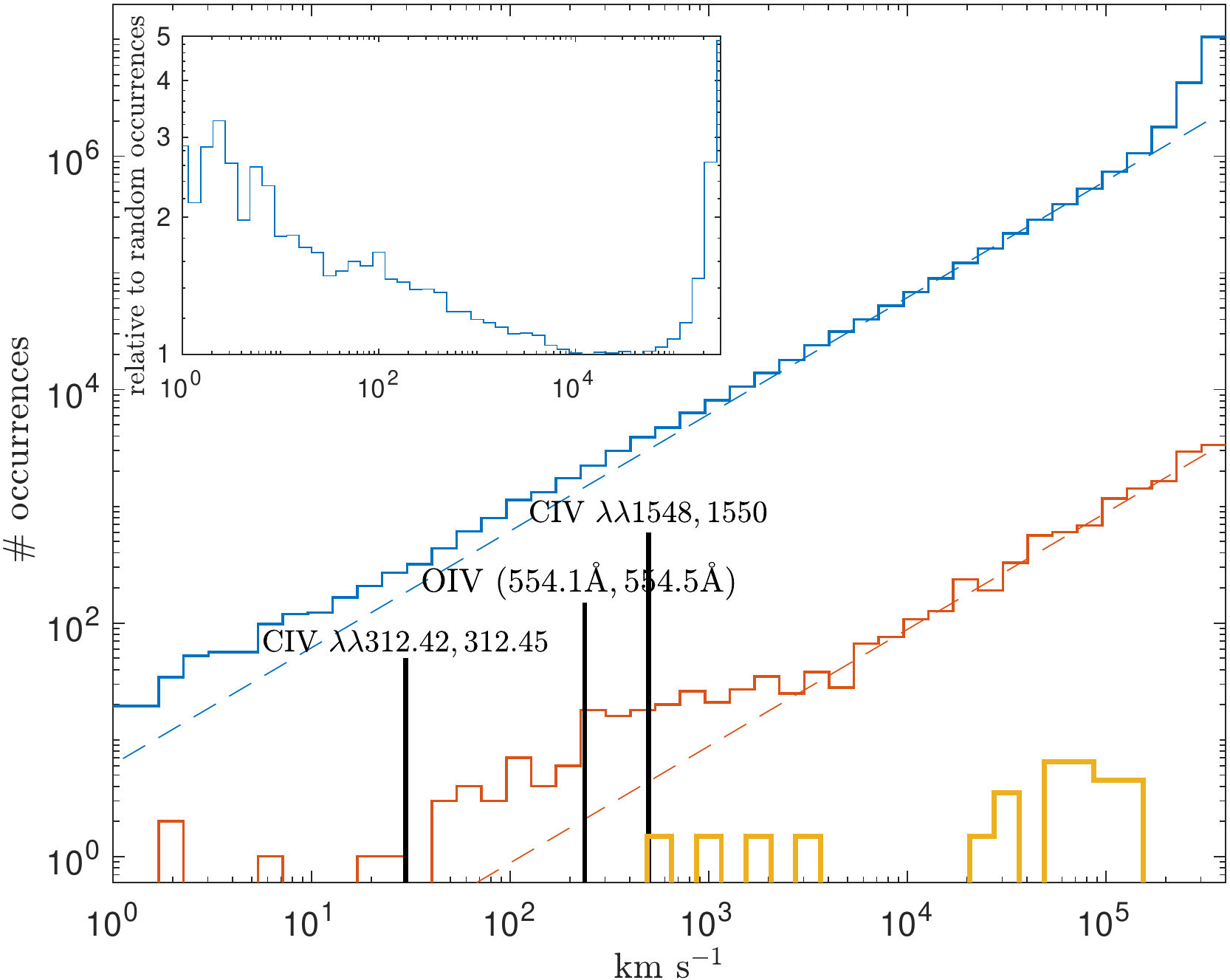}
\caption{Velocity difference, $\delta v$ distributions for resonance lines. The blue curve shows $\delta v$-distribution for all transitions with an oscillator strength greater than 0.1. Deviations from a random distribution (dashed blue line) are seen at the low velocity end due to atomic physics, and at the high-velocity end due to special relativistic effects (see inset). Focusing only on transitions relevant to highly-ionized gas (see text) leads to a similar distribution (red line). In both cases, there are transitions which are theoretically able to LL at $\delta v<500\,\mathrm{km~s^{-1}}$. Introducing a spectral cutoff beyond the Lyman edge results in \ion{C}{4}\,$\lambda\lambda 1548.19,1550.77$ being the first strong transition to LL. \label{LLS}}
\end{figure}

Theoretically, continuum absorption by resonance transitions drives the gas at the implied densities with the contribution of absorption from excited levels being negligible. The wavelength distribution of resonance lines leads to a spectrum of velocity differences, which deviates from a random distribution due to atomic physics at velocity-separations $\lesssim 10^4\,\mathrm{km~s^{-1}}$  (Fig.~\ref{LLS}). Focusing on a subset of atomic transitions, which is relevant for optically thin $U\sim 1$ gas\footnote{Here we take all transitions with oscillator strengths $>0.1$ for all prominent ions of all abundant elements for which the ionization fraction is $>0.1$, for a total of 167 transitions.}, shows a similar behavior. Notably, the velocity difference of the CIV\,$\mathrm{\lambda\lambda 1548,1550}$ doublet is not expected to be the lowest velocity difference to line-lock. Below we aim to qualitatively address this issue for a few particular examples.

The CIV\,$\mathrm{\lambda\lambda 312.42,312.45}$ has similar oscillator strengths to the CIV\,$\mathrm{\lambda\lambda 1548.19,1550.77}$, so that their contributions to the radiation pressure force can be significant if the monochromatic FUV luminosity dominates. Specifically, for clouds whose initial velocity separation is negligibly small, one expects the systems to lock first at a velocity separation of $\sim 30\,\mathrm{km~s^{-1}}$. Realistically, however, the troughs widths are comparable to the LL velocity difference \citep{ham11}, hence random motions in the medium could mask out LL signatures or even prevent locking from occurring at such subsonic speeds.

The OIV multiplet has its $554.1\text{\AA},~554.5\text{\AA}$ transitions separated by $\simeq 237\,\mathrm{km~s^{-1}}$, and oscillator strengths which make their contribution to the radiation pressure comparable to that of the CIV doublet. Such a velocity difference between line-locked systems has not been statistically uncovered by large surveys \citep{mas19}, nor detected in large numbers in high-resolution data. This could imply that gas conditions -- whether composition or ionization state -- are less conducive to LL by OIV, or that the quasar SED is much softer than assumed here, perhaps due to continuum absorption shortward of the Lyman edge. Alternatively, the clouds may have an initial velocity difference that exceeds $\simeq 237\,\mathrm{km~s^{-1}}$ once accelerated along our sightline to the quasar.

\begin{figure*}[t!]
\epsscale{1.18}
\plotone{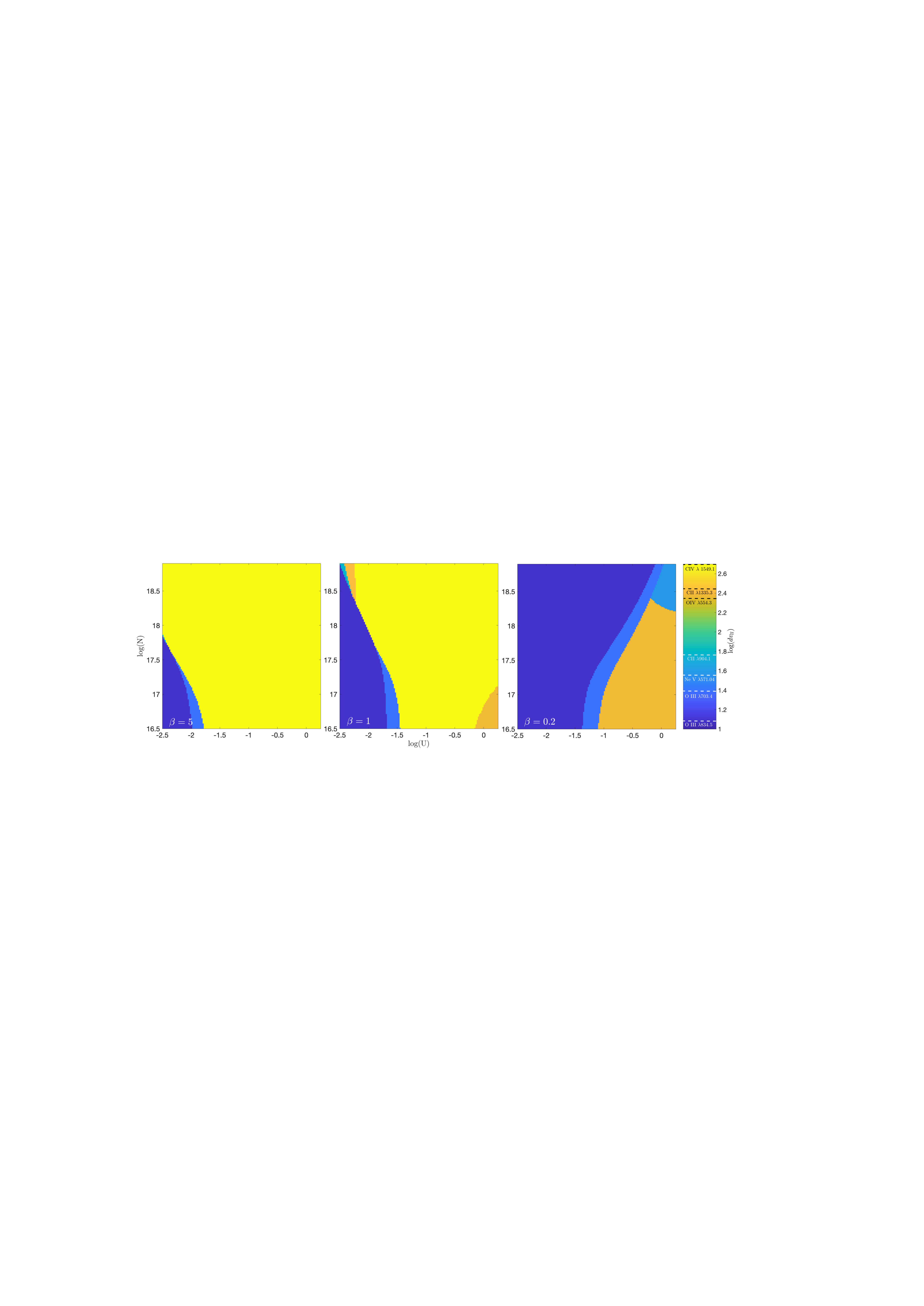}
\caption{The estimated minimal line-locking for clouds of given identical properties, which remain fixed over the outflow dynamical timescales. In each panel a different set of absorption lines is considered, which satisfies Eq.~\ref{beta} for different values of $\beta$ {\it Left:} $\beta =5$ results showing that locking at the \ion{C}{4}\,$\lambda\lambda 1548.19,1550.77$ doublet separation occurs for much of the available phase space. {\it Middle:} results for $\beta =1$ that correspond to the standard model, imply that this simple model is not inconsistent with the observations and the implied gas properties, and that LL at $500\,\mathrm{km~s^{-1}}$ is expected. {\it Right:} results for $\beta =0.2$ predicted LL at lower velocities due to other multiplets (see text). \label{llv}}
\end{figure*}

More generally, for two {\it identical} clouds that are exposed to a flat $\nu L_\nu$ (this is justified to within a factor of two in the range 700\AA--20,000\AA\ for our chosen SED) and following from equation \ref{dMC4}, the ratio between $\delta M$ due to perfect LL by some multiplet $X$ (when the troughs perfectly overlap in velocity space so that $\delta v=\delta v_{ll}$) to that due to the \ion{C}{4}\ doublet is
\begin{equation}
\frac{\delta M_{\delta v_{ll},\mathrm{X}}}{\delta M_{\delta v_{ll},\mathrm{CIV}}}\simeq \left \{ \begin{array}{cc}
  (\tau_\mathrm{X}/\tau_\mathrm{CIV})^2   &  \tau \ll 1\\
  \sqrt{\mathrm{ln}(\tau_\mathrm{CIV})/\mathrm{ln}(\tau_\mathrm{X})}   & \tau \gg 1
\end{array} \right.,
\end{equation}
where the optical depth ($\tau$) limits considered apply to both transitions. The limit $\tau \gg 1$ is included for completeness and is less relevant as the contribution of very optically thick lines to $M$ is small (e.g., Fig. \ref{mom}). Here we neglected small thermal broadening differences between different metal lines. We numerically evaluate Eq.~\ref{dMC4}, and show the minimal velocity at which LL is expected to occur based on the following prescription: for each model defined by $[U,N]$, all multiplets that satisfy  
\begin{equation}
\frac{\delta M_{\delta v_{ll},\mathrm{X}}}{\delta M_{\delta v_{ll},\mathrm{CIV}}}\ge \beta, 
\label{beta}
\end{equation}
are included, and the minimal velocity at which LL occurs is associated with the multiplet having the smallest velocity separation. We first consider $\beta=1$ as threshold with the underlying premise being that transitions with $\delta M_{\delta v_{ll},\mathrm{X}}$-values comparable to or larger than that of the \ion{C}{4}\ doublet are more likely to line-lock first if their $\delta v_{ll}<500\,\mathrm{km~s^{-1}}$ given the larger phase-space volumes associated with them\footnote{For (nearly) identical and co-spatial clouds, LL will occur at the minimal velocity separation between transitions, which is just above the effective thermal speed of the medium, and once the kinematic effects of pressure gradients subside.}. The calculations imply that for much of the phase-space that appears to be relevant to line-locked systems \citep{ham11,bow14} the \ion{C}{4}\ doublet is more likely to lock first (see Fig.~\ref{llv}). This is not the case, however, for low ionization systems with $U\lesssim 10^{-2}$, which may lock at velocities of order the line broadening due to \ion{O}{3} transitions. High ionization ($U>1$) low column systems are more likely to lock at velocities that correspond to that of \ion{O}{4}\ transitions. 

To qualitatively assess the degree to which composition and/or changes to the quasar SED could affect LL velocities by different transitions we resort to a very simplified prescription whereby only transitions that satisfy Eq.~\ref{beta} with $\beta \ne 1$ are included. In case no transitions are found that satisfy the criterion, or those that do satisfy it imply a $\delta v_{ll}>500\,\mathrm{km~s^{-1}}$ then the LL velocity is set to $500\,\mathrm{km~s^{-1}}$. For $\beta=5$, this approach effectively boosts the relevance of the CIV\,$\mathrm{\lambda\lambda 1548,1550}$, which could be due to a relative suppression of the EUV flux of the quasar -- perhaps due to continuum shielding -- or due to enhanced carbon abundance in the gas (see \S5.2). Under this criterion, the phase space leading to $\delta v_{ll}\simeq 500\,\mathrm{km~s^{-1}}$ is enlarged, covering much of the relevant $[U,N]$ plane (Fig.~\ref{llv}). Changing the criterion to $\beta = 0.2$ includes many more transitions, which are able to LL at lower velocities (perhaps due to enhanced EUV flux or a reduced carbon abundance), and at no point in phase space does the system lock at $\delta v_{ll}>240\,\mathrm{km~s^{-1}}$. Specifically, systems with properties similar to those observed in J\,2123-005 are then more likely to lock by the aforementioned \ion{O}{4}\ doublet. The fact that many LL systems are identified at $\delta v_{ll}\simeq 500\,\mathrm{km~s^{-1}}$ implies that the EUV hump cannot be significantly underestimated by our model, or that some mechanism exists, which forms clouds with an initial relative velocity which significantly exceeds the thermal speed, by as much as an order of magnitude before being accelerated along our sightline to the quasar. 

A more quantitative follow-up of LL kinematics involving all candidate transition for LL, which includes the effect of quasar variability and the potential hopping between different line-locked transitions, is beyond the scope of this work.

\section{Discussion}
\label{sec:dis}

Our results imply that the fractional phase-space volume conducive to LL in NAL systems is of order a per-cent or less, and therefore much smaller than implied by recent statistical studies of such systems \citep{bow14}. This suggests fine-tuning of the clouds properties, which sets stringent constraints on their formation path, their evolution over dynamical timescales, and their environment. 

\subsection{Implications for cloud-formation scenarios}

Below we consider a non-exhaustive set of models for the formation of outflowing absorption-line systems in quasars.

\subsubsection{Velocity condensations}

It has been previously suggested that NALs are formed by condensations in velocity space of numerous optically-thin cloudlets spread in velocity space, which undergo line-locking, and accumulate at particular velocities \citep{mil26,sca70,sca73}. Therefore, the formation of line-locked systems in the context explored here is just one manifestation of a potentially more general phenomanon. Our calculations have shown that should cloudlets be formed having a range of densities and column-densities, and with \ion{C}{4}\ significantly contributing to the radiation pressure force, only a very small fraction of those clouds, of order per-cent at most, would be able to line-lock. Further, the more optically thin the clouds are, the higher the degree of fine-tuning required for them to lock since $a_2-a_1 <\delta a_\mathrm{rad} \propto \tau^2 \to 0$ at low opacity (Eq.~\ref{dMC4} and related text in \S4). Such a scenario suggests then that the majority of the material should remain spread out in velocity space, and give rise to very shallow troughs. In that case, however, much higher values of reddening, at the level of $E(B-V)\sim 1$\,mag, would be observed for dust-to-metals ratio typical of the local ISM, and contrary to observations  \citep{bow14}. Further, if numerous cloudlets line lock to make up discrete absorption components then the numbers of high-multiplicity systems will exceed those observed. We therefore consider this scenario unlikely. 

\subsubsection{Turbulent media}

It is intriguing that current density and location estimates for LL systems (as part of the more general NAL population) imply the presence of spatially compact and dense ($\gtrsim 10^3\,\mathrm{cm^{-3}}$) dusty clouds on hundreds of pc scales away from the central black hole. In non-active galaxies, such properties characterize molecular clouds. The formation of molecular clouds, and in particular their cores, is believed to arise from super-sonic turbulence. In this scenario, significant compression occurs due to strong shocks, in which the post-shocked compressed gas can significantly cool and condense. Here we assume that NAL systems are relics of molecular clouds, and follow the statistical properties of a turbulent medium from which they formed. The degree to which this assumption is realistic for an accelerated medium is unclear. 

Recent simulations of supersonic (isothermal) turbulence suggest that the density distribution is of the log-normal type, $P(\rho)\propto \mathrm{exp}\left [ -\left ( \mathrm{ln}\,\rho - \left < \mathrm{ln}\,\rho \right > \right )^2/2\tilde{\sigma}^2\right ]$ over four orders of magnitude in (normlized) density \citep{kri07}. Numerical studies find that $\left < \mathrm{ln}\,\rho \right >=-\tilde{\sigma}^2/2$, where $\tilde{\sigma}^2= \mathrm{ln} \left (1+b^2\mathcal{M}^2 \right)$ with $b\lesssim 1$, and $\mathcal{M}$ is the Mach number. With our density estimates relative to the mean ISM density implying $\mathrm{ln}\rho \sim 8$, the column density distribution, $N (\rho) \sim \rho P(\rho)$, lies on the decaying tail, such that $N(\rho) \propto \rho\, \mathrm{exp}\left [ - \left ( \mathrm{ln} \rho \right )^2/2 \right ]\sim \rho^{1-\mathrm{ln}(\rho)/2}\sim \rho^{-3}$ (we assume $\tilde{\sigma}^2\sim 1$ due to the logarithmic dependence on $\mathcal{M}$, and $\mathcal{M}<10$; \citealt{tof11}). The deduced $N(\rho)$ is very different from the one required for LL to operate, for which $N(\rho) \propto \rho^{-\eta}$ with $\eta<0$ (see Fig.~\ref{ss} and \S3.2.3). It is therefore unlikely that LL systems originate from a turbulent ISM structure that is typical of (non-active) galaxies.

\subsubsection{Mechanically compressed and pushed ISM clouds}

Recently proposed models for quasar outflows suggest that absorption line systems (including BALs) result from the compression and mechanical acceleration of the ISM on galactic scales by a fast and hot wind emanating from the active nucleus \citep{fau12,zei20}. This class of models often does not directly include the effect of radiation pressure force in lines on the cloud kinematics, but can be used to test whether the resulting condensations' properties are consistent with those required by LL arguments, which signify the dynamical importance of radiation pressure force. Specifically, \citet[][see their Fig.~8]{zei20} give predictions for the column density distribution of the compressed ISM clouds, whose velocity-dependent average declines with velocity by $\sim 0.5$\,dex over line-locking velocity separations. Under such conditions, LL is unlikely to materialize. Further, the statistics reported by \citet{zei20} implies that the probability for two clouds to have a velocity separation of $<50\,\mathrm{km~s^{-1}}$ ($400-600\,\mathrm{km~s^{-1}}$) {\it and} have their column densities similar to within 0.1\,dex is $\sim 13$\% ($\sim 10$\%), hence significantly lower than the observed LL statistics. The above probability estimates from the simulations are very likely over-estimated since a large range of gas temperatures -- i.e., densities -- was assumed by \citet{zei20} to provide column-density predictions, while, as our calculations imply, fine tuning of the clouds' density {\it and} column-density is required. Nevertheless, it must be emphasize that the relevance of the \citet{zei20} simulations to high-velocity NALs has yet to be worked out since the velocity range, and column density range included in those simulations is different than the observed ones, and relevant radiation pressure force terms not included in their work.

\subsubsection{Medium instabilities}

Perhaps the greatest challenge of (radiation-) hydrodynamic instabilities in explaining the emergence of LL in quasars is the high-level of fine-tuning required to facilitate LL between  physical components of the outflowing medium. Therefore, drawing robust conclusions requires detailed numerical simulations, which are unavailable for the problem at hand. Here we make no attempt to do so, and resort instead to qualitative analytic arguments for a few cases of interest.

Thermal instability is a plausible means to form discrete entities -- ''clouds" -- of cool  condensations from a more dilute and hot medium \citep{mo96,bra07}. For the chosen SED, the gas is thermally stable under isochoric conditions. Further, LL optimally occurs for ionization parameters which are also thermally stable under isobaric conditions. Marginal stability, but not formal instability, exists for $10^{0.8}<U<10^2$ in our model, so that gas components that cover the temperature range $3\times 10^4-2\times 10^5$\,K may be in pressure equilibrium. Therefore, for the particular model explored here, thermal instability is unlikely to give rise to the observed condensations, and to the narrow range of systems properties implied by LL considerations.

Consider also a more dynamical scenario in which thermally unstable gas under isobaric conditions is exposed to a varying quasar flux with period $t_\mathrm{var}$ (we neglect other perturbations in our highly simplified description). In this case, gas whose cooling/heating timescales are short, as is in our case, could settle to a stable thermal state if isobaric conditions are achieved in regions whose sound-crossing timescale satisfies, $t_s\sim (N/n)/c_s\sim t_\mathrm{var}$, where $c_s$ is the sound speed in the hot medium with temperature $T_h$, and $N$ is the column density. This leads to the condensations' column densities satisfying $N(t_\mathrm{var})\propto t_\mathrm{var}$. As quasars vary over a range of timescales, the column density distribution is not single valued, and would probably be broadened by a myriad of additional processes not included here. While the true effect must be calculated numerically, it is unclear why a high degree of fine-tuning of the clouds properties may be provided by such a process.

A further challenge for this model concerns reddening constraints. Specifically, the column density of the volume filling hot medium from which the cool gas condenses, and with which it is in pressure equilibrium, is given by $r_0n_c(T_c/T_h)\sim 10^{21}r_\mathrm{100pc}\,\mathrm{cm^{-2}}$, where we assumed $n_c\sim 10^3\,\mathrm{cm^{-3}}$ and $T_h\sim 10^6$\,K. This implies significant rest-frame visual extinctions of $\sim 0.5$\,mag toward quasars, which is not observed. Thus, unless dust-formation occurs in-situ, thermal instability is an unlikely origin for high-velocity LL NALs. 

Another type of instability is the line-driven instability (LDI), which is thought to operate in the winds of massive stars \citep{owo84}, and has been suggested as a possible scenario in the context of LL systems in quasars \citep{bow14}. In this scenario, small velocity perturbations of the outflow lead to shadowing/de-shadowing effects among its different parts. These cause non-monotonic spatial variations in the radiation-pressure force, which lead to non-monotonic velocity fluctuations in the flow, hence to growing density stratification and to shocks. These result in a multiphase structure with typical length scales of order the Sobolev length scale of the flow, $l_\mathrm{Sob}\simeq \sigma /(dv/dr)$, where $dv/dr$ the (local) velocity gradient \citep{sun18}. Unlike stellar winds, the Sobolev length scale for NAL clouds is of order the entire cloud length, and multiple NAL-systems statistic do not support a stellar-wind--like scenario. Further, considering published numerical calculations of LDI, it is not clear that the level of fine-tuning, which is required for LL to operate, may be reached \citep{sun18}. Further, we expect LDI to be less prominent in NALs, which are primarily driven by continuum processes (light absorption by dust) and are sensitive to the flow ionization level rather than merely to the line-opacity between different phases of the flow. Thus, it is unclear how relevant LDI is to NAL flows in quasars.

It may be interesting to examine the development of Rayleigh-Taylor instability (RTI) at the  leading (non-illuminated) face of a radiatively accelerated NAL cloud through a dilute ambient medium. In the self-similar phase of RTI (assuming one develops within $t_\mathrm{dyn}$), the mixing layer between the dense and dilute media expands with time, $t$, such that its time-dependent scale-height $h(t)\sim a_{\rm rad} t^2$, where we assume a high-density contrast between the phases (Atwood number of order unity) and incompressibility which are clear over-simplifications \citep{ris04}. At its leading edge, the mixing layer therefore expands at a speed of $v\sim a_{\rm rad}t$, and material -- hereafter extrusions -- whose velocity exceeds the thermal speed to the cloud, will be de-shadow, and hence able to accelerate more efficiently and extrude to ultimately detach from the parent cloud. The column density of the extrusion is estimated here by the product of the density of the parent cloud medium (the degree to which this holds in reality needs to be verified by appropriate simulations) and the scale-height where de-shadowing occurs, $h\sim  (a_\mathrm{rad}/2)(\sigma/a_\mathrm{rad})^2$ gives
\begin{equation}
    N\sim \frac{1}{M\sigma_T} \left (\frac{\mathcal{U}_{\rm gas}}{\mathcal{U}_{\rm rad}} \right )\sim 10^{18}\frac{n_{4}T_4r_{\rm 100pc}^2}{L_{48}M_3} \,{\rm cm^{-2}},
\end{equation}
where $\mathcal{U}_{\rm gas}$ ($\mathcal{U}_{\rm rad}$) is the gas- (radiation-) energy density. Such columns are in the rough ballpark of the column densities found by \citet{ham11}, but it remains to be seen whether such a mechanism can consistently operate and lead to the fine-tuning required for LL. 

\subsubsection{Radiation-pressure confined clouds}

It has been shown that radiation pressure confined (RPC)  gas can achieve remarkably uniform structure regardless of the initial/boundary conditions imposed \citep{bas14,ste14}, and hence is a promising candidate for producing clouds whose properties are highly correlated, as required by LL conditions.

The radiation-to-gas pressure ratio is given by
\begin{equation}
    \frac{P_\mathrm{rad}}{P_\mathrm{gas}}= \frac{\tau_e M}{n k_\mathrm{B}T}\frac{L}{4\pi r^2 c}\simeq 10^{-2}N_{17} U^\gamma,
\end{equation}
where the Compton optical depth, $\tau_e$ is assumed to be $\ll M^{-1}$, and the powerlaw index satisfies $\gamma\simeq 0.7$, which incorporates the dependence of the total force multiplier on $U$ and of the gas temperature on $U$ in the range $10^{-2}<U<10$ (not shown). Here $N_{17}\equiv N/10^{17}\,\mathrm{cm^{-2}}$, and we neglected the modest dependence of $M$ on the column density for marginally optically thick media. Therefore, clouds whose properties are optimal for LL are characterized by $P_\mathrm{rad}/{P_\mathrm{gas}} \ll 1$ and thus do not provide an indication for a RPC dynamics. Taking into account the observed total columns per system, and the supra-thermal line-broadening does not appreciably change our conclusions (\S3). Furthermore, steady-state RPC requires that the leading edge of the cloud is extremely optically thick (so that the bulk acceleration is negligible, as in the case of BLR clouds) or that there is ram pressure from the ambient medium, which balances the pressure by the compressed gas. The latter scenario, which may be relevant for NAL systems at large \citep{ste14} encounters great difficulties in the context of LL since is necessitates extreme fine tuning between  disparate physical mechanisms being the radiation pressure force and the drag force. We therefore find the RPC scenario to be an unlikely explanation for NAL systems undergoing LL.

\subsubsection{Circumstellar AGB shells}

\begin{figure}[t!]
\epsscale{1.18}
\plotone{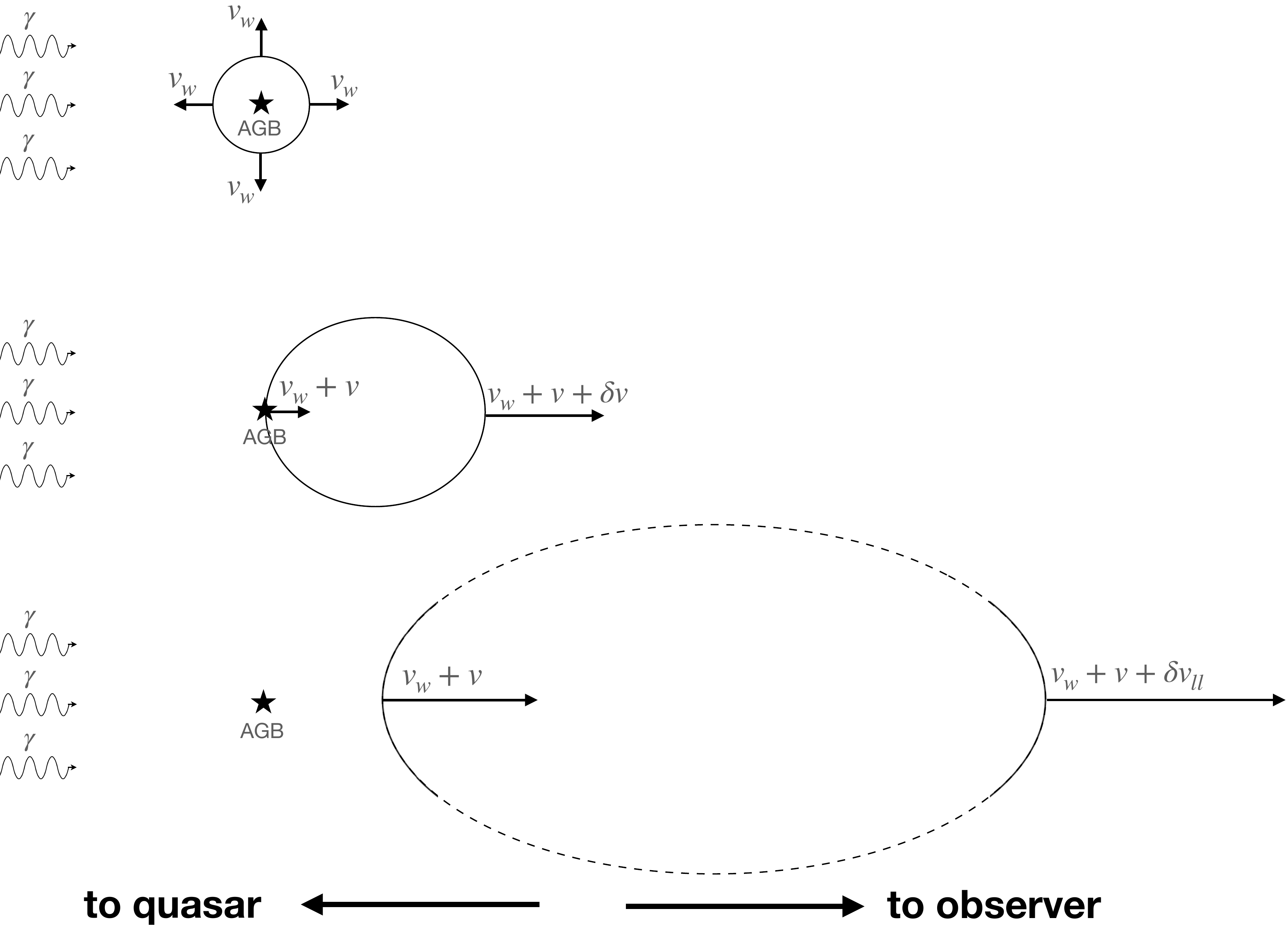}
\caption{A possible depiction of a circumstellar AGB shell evolution toward a line-locked system configuration. A compact geometrically thin shell is ejected during a thermal pulse of the AGB stars (top panel) and expands to gradually cover the quasar continuum emission region while being accelerated by radiation pressure force due to illumination by the quasar (ISM interaction is neglected, perhaps due to ISM pre-evacuation by the quasar). The shell detaches from its origin, accelerates radially from the bulge with its leading edge developing an increasing velocity difference with respect to the trailing side, thereby stretching the shell along the radial direction to the quasar (middle panel). LL velocities are attained over dynamical timescales between the leading and trailing shell rims (lower panel). These drawings are qualitative at best, and the actual nebular shape need not be an ellipsoid (this uncertainty is denoted by dashed shape lines at long timescales). This may be particularly true at locations whose normal vector to the surface is perpendicular to the radial direction (to the quasar), which result in significantly different acceleration of the rims due to optical depth effects. \label{agb}}
\end{figure}

The notion that some quasar outflows originate in continuous stellar winds and their contrails has been suggested by \citet{sco95}. Here we qualitatively examine whether the large expanding circmustellar shells detected, for example, around many (carbon-rich) AGB stars \citep{hof18} could potentially be the origin of LL systems. The model has several appealing attributes: a) it identifies an origin for dense, metal-rich and dusty gas components on galaxy bulge scales, b) it naturally fine-tunes the properties of the two seemingly distinct kinematic components by associating them with a common symmetric origin (a star), and c) it provides an initial velocity separation between the two kinematic components, which are identified with the approaching and receding sides of an expanding shell, thereby preventing LL of nearly identical systems at  relative subsonic speeds (Fig.~\ref{agb}). In addition, it provides a natural explanation for the extremely high aspect ratio implied for some systems \citep{ham11}.

We emphasize that it is not our intention to quantify the level of symmetry that is required by AGB shells to facilitate LL and compare it to available data for nearby AGB shells, nor do we aim to carry out detailed hydrodynamic calculations to study the stability of an expanding shell configuration over dynamical timescales. It is also not within our scope to provide detailed spectral predictions for the absorption and extinction signatures across the electromagnetic spectrum from an ensemble of AGB shells along our sightline to continuum region(s) in quasars.

Here we consider a qualitative model in which the outflow originates in the host galaxy bulge, whose size at $z\sim 2$ is $r_b\simeq 10^{21}M_{b,10}^{1/2}$\,cm \citep{she03,bru14}, where the bulge mass, $M_b=10^{10}M_{b,10}\,\mathrm{M_\odot}$. We parameterize the launching radius of the outflow, $r_0=\epsilon_r r_b$, where $\epsilon_r \lesssim 1$ for a bulge origin of the outflow. In this case, the dynamical timescale of the NAL outflow satisfies $t_\mathrm{dyn}\sim 10^4\epsilon_r r_{b,21}v_{\infty,4}^{-1}$\,years during which time it should be detectable as a NAL system, namely it should substantially cover the continuum emission region of the quasar and have a non-negligible optical depth in relevant UV transitions. 

Geometrically, the radius of the expanding AGB wind over dynamical timescales, neglecting ISM interaction, is 
\begin{align}
    r_{w}\sim \epsilon_r r_b\frac{v_w}{v_\infty} & \simeq 10^{18}\epsilon_r r_{b,21}v_{w,10}v_{\infty,4}^{-1}\,\mathrm{cm}, \label{erw}
\\
\nonumber    & \simeq 10^{19}\epsilon_r^{3/2} \Gamma_\mathrm{Edd}^{-3/4}L_{48}^{1/4}v_{w,10}\,\mathrm{cm} 
\end{align} 
where the AGB wind speed $v_w=10v_{w,10}\,\mathrm{km~^{-1}}$. In the last step we used Eq.~\ref{vinf} (with $M_3=3$) and assumed a bulge-BH-mass relation such that $M_b\sim 100M_\mathrm{BH}$ \citep[but note that this does not apply to pseudo-bulge and pure-disk systems; \citealt{kor11}]{har04,pen06a,din20}, and recast the expression in terms of the Eddington ratio of the quasar, $\Gamma_\mathrm{Edd}$. Clearly, if the current quasar luminosity does not reflect on the average luminosity over the dynamical time, or the gas acceleration changes markedly with distance aside from geometrical flux-dilution effects then the last step may lead to erroneous conclusions about $r_w$. Over dynamical timescales, the AGB wind could therefore fully cover the accretion disk having a half-light radius of $r_\mathrm{SS73} (\lambda) \sim 10^{16}L_{48}^{1/2}\lambda_{1550}^{4/3}$ (the inner disk boundary is ignored here), but may only partly cover the broad-line-region whose size, $r_\mathrm{BLR}\simeq 3\times 10^{18}L_{48}^{1/2}$\,cm (an optical to bolometric luminosity correction of 10 was assumed for the BLR size-luminosity relation of \citealt{ben13}). This could give rise to partial-coverage effects in the absorption troughs due to the finite contribution of the BLR to the continuum signal \citep[][and references therein]{chel19}.

Interestingly, the radial gap between the rims of the expanding shell $\delta r$ satisfies $\delta r/r_0\sim \delta v_{ll}/v_\infty\sim 5\times 10^{-2} \lesssim \vert \delta M_{\delta v_{ll}} \vert /M$ for optical depth in the CIV transition of order unity (Fig.~\ref{dM}), thereby facilitating LL despite the distance gap developing between the kinematic components over dynamical timescales. This is especially true for high metalicity, but dust poor gas.

To absorb in the UV, the wind's ionization parameter should be of order unity \citep{ham11,bow14} over dynamical timescales hence on size-scales of order $r_{w}$. For the chosen SED the following relation holds: $U\simeq L_{48}n_4^{-1}r_{b,21}^{-2}$. This sets a requirement on the ''instantaneous" mass loss rate that leads to the expanding shell of 
\begin{equation}
    \dot{M}\sim 4\pi \rho \epsilon_r^2 r_b^2 \frac{v_w^3}{v_\infty^2}\sim 6\times 10^{-3} \frac{\epsilon_rv_{w,10}^3}{U \Gamma_\mathrm{Edd}^{1/2}}L_{48}^{1/2} \mathrm{M_\odot~yr^{-1}}.
    \label{edotm}
\end{equation}
In comparison, the maximal momentum driven mass-loss rate that can be propelled by a star of luminosity $L_\star\lesssim 10^5L_\odot$ \citep{ven18} is $\dot{M}_\mathrm{max}\sim L_\star/cv_w\sim 10^{-4}\,\mathrm{M_\odot~yr^{-1}}$ (this limit can increase by a factor of a few when multiple photon scatterings in an optically thick non-porous media is involved). Unless the AGB ejecta on large scales are characterized by $v_{w,10} \ll  1$ (perhaps due to mass loading from the ISM) then the model favors a more efficient radiative acceleration, such as due to a lower gas-to-dust ratio than the value adopted here of $\simeq 100$ (see however \citealt{mae18}), or due to a more compact launching region around the quasar than assumed here (e.g., $\epsilon_r\lesssim 0.1$), or their combination. 

Next we estimate the column density through the expanding shell at $r_w$. Denoting the timescale for the thermal pulse during which the shell is ejected by $t_\mathrm{tp}$, then the shell thickness is $\delta r\sim v_w t_\mathrm{tp}$, which remains constant during its expansion due to mass conservation \citep[we neglect mass loading by ISM or precursor wind interaction; ][]{mat07}. Denoting the ejected shell mass by $M_s$ where $M_\mathrm{s}=\dot{M} t_\mathrm{tp}$, then $N\sim n\delta r$, which satisfies 
\begin{equation}
    N\sim \frac{M_\mathrm{s}}{4\pi m_p r_b^2}\frac{v_\infty^2}{v_w^2}\sim 6\times 10^{14}\epsilon_r^{-3}\Gamma_\mathrm{Edd}^{3/2}L_{48}^{-1/2}M_\mathrm{s,-3}\,\mathrm{cm^{-2}},
    \label{eN}
\end{equation} 
where $M_\mathrm{s}=10^{-3}M_\mathrm{s,-3}\,\mathrm{M_\odot}$ with $M_\mathrm{s,-3}\lesssim 10$ is typical of detached AGB shells \citep{olo96}. To match the observed columns of \citep[$\sim 10^{19}\,\mathrm{cm^{-2}}$; ][]{gan03,ham11} the model (again) favors more compact launching regions satisfying $\epsilon_r\lesssim 0.1$. Alternatively, the observed column might result from the confluence of many low-column systems. Nevertheless, this requires all of them to be fine-tuned to yield LL, which is highly improbable (see \S5.1.1). 

The global covering fraction of AGB shells over the quasar sky, $C_g\sim n_\mathrm{AGB}(t_\mathrm{dyn}/t_\mathrm{AGB})r_w^2r_b$, where we assumed that AGB shells survive for a dynamical time, and that all AGBs go through a thermal-pulse phase during their AGB-phase lifetime, $t_\mathrm{AGB}$, whereby a single detached shell is ejected (see, however, \citealt{kas21} for a discussion of multiple shell ejection events from AGB stars with periods of $\gtrsim 10^4$\,years). Noting that carbon-rich AGB are solar-like stars, we estimate $n_\mathrm{AGB}=n_\star(t_\mathrm{AGB}/t_\star)$, where $t_\star$ is the lifetime on the main sequence and with $n_\star=\epsilon_b M_b/({M_\star}4\pi r_b^3/3)$. Here $M_\star$ is the typical stellar mass assumed without loss of generality to be solar and hence $t_\star\simeq 10^{10}$\,years.\footnote{ The above estimates depend little on the assumed $M_\star$ normalization, for a given stellar population, since the number of stars at mass $M_\star$ is $\propto M_\star^{-2.35}$ for a Salpeter initial mass function, while their lifetime is $\propto M_\star^{-2.5}$.} The parameter $\epsilon_b$ is the fraction of the mass bulge that is relevant for producing LL signatures (note that $\epsilon_r$ and $\epsilon_b$ are inter-dependent parameters via the density profile of the bulge; see below). With these definitions
\begin{equation}
    C_g\sim \frac{\epsilon_b M_b}{M_\star} \frac{t_\mathrm{dyn}}{t_\star} \left ( \frac{v_w}{v_\infty} \right)^2 \sim 50 \epsilon_b\epsilon_r^{3/2}L_{48}^{3/4}\Gamma_\mathrm{Edd}^{-9/4}v_{w,10}^2.
\label{cg}
\end{equation}
Taking $\epsilon_r=0.1$ and $\epsilon_b=0.01$, results in $C_g\sim 0.01$, which is of order the observed value \citep{che21}. Our choice of $\epsilon_b$ is consistent with the presence of a compact nuclear star cluster \citep[NSC,][]{neu20}, whose mass is $\sim 10^{-3}-10^{-2}M_b$ in local sources \citep[and references therein]{geo16}\footnote{We emphasize that the outflow scenario advocated here may operate in addition to other mechanisms that my lead to outflow phenomena from NSCs \citep{goh18}.}. Our $C_g$ estimate does not account for time-dependent star-formation history, and depends on the survival time of accelerated AGB shells, as well as on the number of shells ejected during the AGB lifetime. A more realistic estimation awaits numerical simulations, which are beyond the scope of this work, and a more comprehensive comparison between model predictions and absorption signatures in the UV and X-ray range over the luminosity range that characterizes active galactic nuclei.

In the above model an expanding shell from an AGB star, which is driven along its sightline to the quasar, will dislocate with respect to its origin and eventually detach from its parent star (middle panel of Fig.~\ref{agb}). The shell radius at detachment, $r_d$, is crudely given by
\begin{equation}
    r_{d}\sim 2\epsilon_rr_b\left ( \frac{v_w}{v_\infty} \right )^2\sim  10^{15}\epsilon_r\Gamma_\mathrm{Edd}^{-1}\,\mathrm{cm}. 
\label{detach}
\end{equation}
For our adopted formalism to be consistent, we therefore require that $r_{d}>R_\mathrm{AGB}$, which is the photospheric radius of a typical AGB star, which we take to be $100R_\odot$ \citep{hof18}. 

\begin{figure}[t!]
\epsscale{1.18}
\plotone{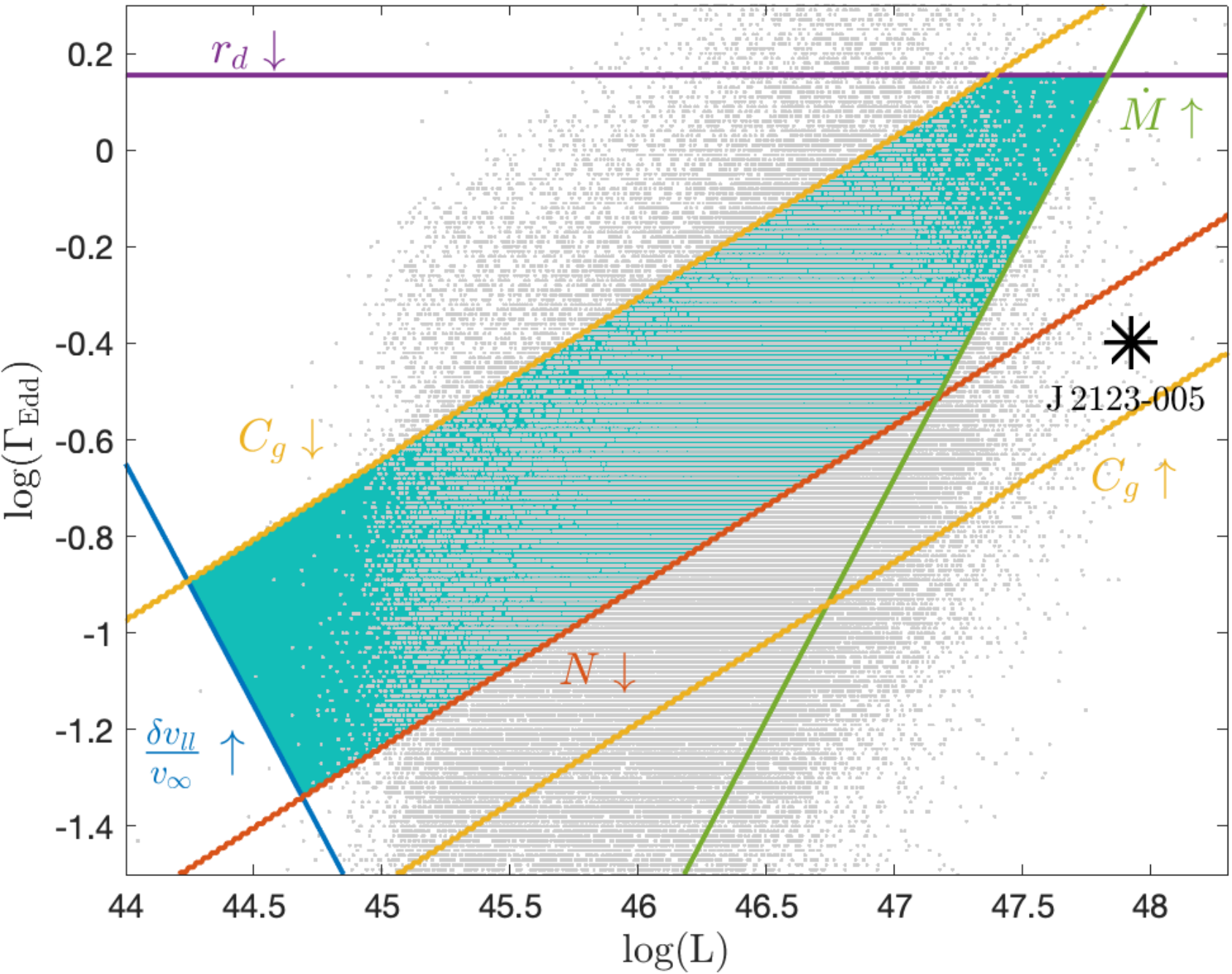}
\caption{The allowed phase space for LL systems within the AGB-shell model. The colored fill patch is the allowed phase space of the model, and is determined by the observational properties of LL systems, and known circumstellar shells around (local) AGB stars (se text). Different boundaries delineating the allowed phase space are set by different conditions (see color-coding) with the relevant quantity and its gradient away from the colored region denoted next to each curve. The black symbol marks the phase-space position of J\,2123-0050. We emphasize that different assumption about $\epsilon_r,~\epsilon_b,~M_s$ and $v_w$ can substantially expand or shrink (or even void) the phase space available for LL. Gray points are quasars from the \citet{shen11} sample, demonstrating that many of them lie within the allowed phase-space of the particular model shown. \label{allow}}
\end{figure}

We next provide a first stab at mapping the quasar phase space where LL could occur according to this model. Motivated by the above considerations for the more relevant range of parameter values we use $\epsilon_r=0.03,~\epsilon_b=0.02,~v_{w,10}=2,~M_{s,-3}=10$, and $U=1$, and consider the phase space spanned by the remaining parameters, namely $L$ and $\Gamma_\mathrm{Edd}$ (Eqs.~\ref{erw}-\ref{cg}). As for observational constraints, we require that $\dot{M}<10^{-3}\mathrm{M_\odot~yr^{-1}}$ (set by local AGB physics), $\delta v_{ll}/v_\infty <5\times 10^{-2}$ (set by Eq.~\ref{ineq}; see Fig.~\ref{NNUt}), $17<\mathrm{log}(N)<{20}$ \citep{ham11,bow14}, $-2.5<\mathrm{log}(C_g)<-0.5$  \citep{che21}, and $r_w\geq 10^{16}L_{48}^{1/2}\lambda_{1550}^{4/3}$ (full coverage of the UV emitting disk). We also require that $r_d>100R_\odot$ (see above). The allowed phase space is shown in figure~\ref{allow} and includes a substantial fraction of the Sloan digital sky survey (SDSS) sources \citep{shen11}. We emphasize that the phase-space volume is sensitive to $\epsilon_r,~\epsilon_b,~M_s$ and $v_w$, all of which are rather uncertain, and some parameter combinations may void the model altogether. 

Generally, the model predicts that sources at the top range of the Eddinton-rate distribution at a given luminosity bin, are less likely to show LL since the covering fraction is low. In the bottom range of the Eddington-rate distribution, LL systems are characterized by low columns of gas, resulting in weaker absorbers, which may surface with high-resolution spectroscopic surveys, and with higher rate of occurrence per source due to the higher covering fractions implied. The source J2123-005 formally lies outside the phase space predicted by the specific model considered here, which results from the assumed condition on the peak mass-loss rate from AGB stars. Objects in this range may show higher ionization absorption. Lastly, fainter sources emitting at low Eddington rates are not expected to show LL systems unless carbon is highly over-abundant or the gas is dust-poor.

\subsection{Observational tests of the theory and their implication}

Our calculations indicate that the most probable configuration for LL is that of clouds with similar, but not strictly identical properties, such as ionization and column density (and also gas composition). In particular, it is required that $a_2\gtrsim a_1$, for LL to occur. A challenge to this theory would be to find counter examples, namely, that the faster (shielded) component has, for example, a higher column density and a higher ionization level than the low-velocity component, so that $a_2<a_1$ (unless the system is at its coasting phase). 

The most likely phase space for LL by the \ion{C}{4} doublet to occur is for clouds with column densities of order $10^{17}\,\mathrm{cm^{-2}}$. For clouds whose inner velocity dispersion is significantly above the thermal values, the optimal column-density scales with the effective line width, at least so long as the continuum optical depth is smaller than unity. For the case of J\,2123-005, a suprathermal line broadening of $\sim 30\,\mathrm{km~s^{-1}}$ was found, which implies optimal columns for LL of $N\lesssim 10^{18}\,\mathrm{cm^{-2}}$. This is within a factor of a few of the column-density estimates of \citet{ham11}. While higher column clouds can experience LL, their properties need be extremely fine-tuned, which imposes extremely tight constraints on the physical mechanism leading to cloud formation, and controlling cloud stability over their acceleration timescales.

Clouds with very different properties must occupy an extremely localized and fine-tuned range of the phase space to facilitate LL. Further, such line-locked cloud configurations can be easily disrupted by luminosity variations of the center source over dynamical timescales. Quantifying the relation $U_1/U_2$ and $N_1/N_2$ for line-locked NALs, and comparing those to model predictions (e.g., Fig.~\ref{NNUt}) will shed light on the physics of such systems. 

Our calculations indicate that for compositions of order the solar value with ISM-like dust-to-metals mixture, the ratio between the radiation pressure force term giving rise to LL and the total radiation pressure force is order order per-cents, which by dynamical-time arguments implies that the expected $dv_{ll}/v_\infty$ is of the same order. Finding systems for which the latter ratio is much higher -- i.e., low velocity systems experiencing LL -- would imply that the gas composition may be significantly different than assumed, with the abundance of the element giving rise to LL being particularly enhanced. Indeed, \citet{bow14} find evidence for LL in systems with $dv_{ll}/v_\infty\gtrsim 0.17$, which might indicate an overabundance of carbon by an order of magnitude or more compared to the solar composition \citep[see][who considered such models for AGB stars and noted the weak thermal pulses associated with them]{kar22}. Alternatively, it could mean that some systems are relatively dust-poor so that the total radiation pressure force is lower -- by roughly a factor of 10 for $U=1$ and a column of $10^{19}\,\mathrm{cm^{-2}}$ -- and the relative contribution of the line-locked transitions to the total radiative acceleration is correspondingly higher, and hence larger $dv_{ll}/v_\infty$ ratios may be reached, by roughly a factor of 3. Metal-poor massive stars are thought to have low dust-yields \citep{del19}, which could explain the low terminal outflow velocities associated with some LL systems. It will be interesting to examine whether low-velocity LL systems show less reddening per their absorbing columns than high velocity systems.

The kinematic models employed here suggest that quasar variability over dynamical timescales can be disruptive for line-locked systems, which have not yet reached their coasting phase. It would be interesting to check whether LL systems have a higher incidence in less variable sources or use dynamical arguments involving LL to constrain the structure function (power-spectrum) of quasars. Further, looking for $\delta v$-statisics of multiple quasar NALs could be used to check for non-uniform distribution over velocity space, perhaps related to the predicted bifurcation patterns (Fig.~\ref{qvar}). 

It is of considerable interest to search for LL at velocity separations corresponding to multiplets other than \ion{C}{4}\,$\lambda\lambda 1548.19,1550.77$, and assess their probability, which could shed light on the kinematics of NAL systems. For example, finding ample LL systems at small velocity separations whose outflow velocities are moderate, could indicate an evolutionary path by which NAL clouds hop from one line-locked position to the next as their properties change across their path, or due to quasar-flux variability. Conversely, not finding evidence for LL at small velocity separations could mean that clouds are formed with a finite velocity difference between them, ruling out, for example, Rayleigh-Taylor instability as the origin of multiple NAL systems.

The proposed scenario in which circumstellar AGB envelopes may be the origin of LL systems should be further investigated by searching for commonalities between the metal and dust content of LL systems and those of shells around local AGB stars (having in mind the different redshift range probed in each case; \citealt{mar06}), and specifically carbon-rich ones, which likely result from thermal pulses rather than from ejecta-ISM interaction. Further, calculation of the global covering factor by a population of circumstellar AGB shells, which uses realistic AGB population and evolution models, should be confronted with LL statistics. Importantly, numerical simulations must be performed to test for the stability of shells as they are accelerated through the dilute, perhaps pre-evacuated bulge medium in quasar hosts, and realistically assess the probability of LL to occur along quasar sightlines (i.e., of $a_2\gtrsim a_1$). On the flip side, the study of LL systems can resolve small scale phenomenon in  AGB ejecta via partial covering effects of the quasar continuum source. This pencil-beam approach could shed light on the physics and composition of expanding AGB shells, and improve models for such objects in the local universe. It may be interesting to check whether highly supersolar carbon abundances, which may be implied by the existence of systems with large values of  $dv_{ll}/v_\infty$  \citep[][see above]{bow14} may be reconciled with our understanding of local AGB ejecta 

Lastly, the detection of multiple ($>2$) LL systems is a challenge to the simple AGB scenario outlined here, and it remains to be tested whether the spiral circumstellar material patterns seen around local AGB stars due to binary interaction \citep{hof18} could provide a viable explanation for this phenomenon. 

\section{Summary} \label{sec:sum}

The emergence of line-locking (LL) of accelerating and outflowing NAL systems in quasars is studied by means of detailed photoionization and kinematic calculations. It is found that only a very small volume of the relevant phase space is conducive to LL, which appears to be at odds with recent findings for the relatively high  occurrence of this phenomenon in multiple-components NALs. This implies a high-degree of fine-tuning between the properties of apparently distinct absorption components, which sets stringent constraints on their formation scenarios. 

Motivated by available constraints on the ionization and thermal state of such systems, the conditions for LL are examined in detail over the relevant phase space, as well as the stability of such configurations against time variations of the quasar flux. We find that the properties of the line-locked NAL system in J\,2123-005, which is perhaps the best studied system of its kind, seem to be in agreement with the phase space optimal for LL due to the \ion{C}{4}\,$\lambda\lambda 1548.19,1550.77$ doublet, after allowing for supra-thermal line-broadening. Further, the ratio of the LL velocity of $\simeq 500\,\mathrm{km~s^{-1}}$ and the outflow velocity in this source is $\sim 5$\%, and is qualitatively consistent with model predictions for the relative contribution of the \ion{C}{4}\ doublet transitions to the total radiative acceleration assuming solar-like metal composition and dust-to-metals ratio.  Nevertheless, for the clouds to develop a velocity difference leading to LL while being accelerated to their bulk outflow velocity, requires extreme fine-tuning of their properties along their entire path, which occupies a negligibly small fraction of the phase-space volume.

The high-degree of fine-tuning between the properties of LL NALs is surprising, and is inconsistent with most NAL formation scenarios, such as thermal instability, the mechanical compression and pushing of ISM clouds, velocity condensations or ''attractors" due to the aggregated effect of LL, or radiation pressure confined clouds. The high degree of fine-tuning needs to be maintained over dynamical timescales of the flow, and is not merely viewed at the coasting phase, thus implying stable clouds whose properties do not vary dramatically over time, and certainly not independently of each other. This is difficult to materialize if non-radiative force terms (e.g., drag) are important since those require further tuning with respect to additional, independently varying physical processes. This suggests that line-locked NALs occur in extremely dilute environments, which may have been pre-evacuated by the quasar. This, however, has implications for the confinement of NAL systems that travel at highly (and slightly different) supersonic speeds, and yet retain their properties over dynamical times.

A scenario that associates line-locked systems with expanding circumstellar AGB shells in the quasar host is proposed, which naturally leads to finely tuned NAL properties, prevents LL small velocity differences between the clouds, and is qualitatively consistent with the observational constraints for well studied systems. Several predictions of the model are provided, and tests of the theory are outlined. If substantiated as a viable model for LL systems then it could provide a unique probe of {\it individual} stellar phenomenon in the hosts of quasars at high-redshift, which can be used to shed light on their star-formation and metal-enrichment history, and the properties of the ambient interstellar material in quasar hosts. Additionally, LL systems can be used to assess the mass loss rate from individual AGB stars at epochs when the universe was much younger than today, and may provide a unique probe of the metalicity and gas-to-dust mixture in AGB ejecta before substantial mixing occurs. Additional numerical work is required to test the proposed scenario, which is beyond the scope of the present paper.

\begin{acknowledgements}

\noindent This research has been supported by grants from the Israeli Science Foundation, ISF (2398/19), and the German Research Foundation, DFG (CH71-34-3). TRL acknowledges additional support by the Zuckerman Foundation through a Zuckerman Postdoctoral Fellowship, as well as support by the NASA Postdoctoral Program. We thank an anonumous referee for constructive comments and suggestions. We are indebted to P. Goldreich and J. Everett for fruitful discussions in the early stages of this work, and thank M. Zeilig-Hess for helpful feedback. We thank G. Ferland and collaborators for creating and maintaining the {\sc cloudy} photoionization code. Calculations were performed using high-performance computing facilities at the University of Haifa, which are funded in part by an ISF grant (2155/15). 

\end{acknowledgements}

%\newpage 

\bibliographystyle{aasjournal}
\bibliography{BibList}

\end{document}